\documentclass[aps,prd,showkeys,showpacs,amssymb,cite,
amsfonts,epsf,preprintnumbers,nofootinbib,superscriptaddress]{revtex4}

\usepackage[dvips]{graphicx}
\usepackage{bm,latexsym,amsmath,amssymb,amsfonts,color}
\usepackage{soul}

\newcommand{\be}{\begin{equation}}
\newcommand{\ee}{\end{equation}}
\newcommand{\bear}{\begin{eqnarray}}
\newcommand{\eear}{\end{eqnarray}}
\newcommand{\ba}{\begin{array}}
\newcommand{\ea}{\end{array}}

\begin{document}

\title{Numerical Study of Instability of Fluid Black Holes}

\author{Inyong Cho}
\email{iycho@seoultech.ac.kr}
\affiliation{School of Liberal Arts,
Seoul National University of Science and Technology, Seoul 01811, Korea}
\author{Dong-Ho Park}
\email{dongho@kasi.re.kr}
\affiliation{National Fusion Research Institute, Daejon 34133, Korea}

\begin{abstract}
Recently neutral and charged black-hole solutions were found 
for static perfect fluid with  the equation of state $p(r)=-\rho(r)/3$,
for fluid only as well as for fluid in the presence of electric field.
In those works, the stability of the black holes were studied
in an analytic manner, which concluded that the black holes 
are unconditionally unstable.  
In this work, we focus particularly on the {\it numerical} study of the instability.
For the black-hole solutions as well as the static solutions without horizons,
we solve the perturbation equations numerically
and find the unstable mode functions. 
\end{abstract}
\maketitle

\section{Introduction}
Gravitating fluid has been a hot issue for a long time.
Lots of works have been focussed on the time dependent situation
such as the Friedmann universe in the literature,
while gravity of static fluid has also been studied quite intensively
~\cite{Bekenstein:1971ej,Sorkin:1981wd,Pesci:2006sb,Semiz:2008ny,Lake:2002bq,Bronnikov:2008ia,Cho:2016kpf,Cho:2017nhx,Cho:2017vhl,Kim:2017hem,Kim:2019hfp}.
For static fluid, the gravitating solutions have been obtained recently in Refs.~\cite{Cho:2016kpf,Cho:2017vhl}
for the equation of state $p(r) = -\rho(r)/3$.
The purpose of the choice of this equation of state was
to realize the spatial topology of $S_3$ (closed) and $H_3$ (open).
Purely $S_3$ or $H_3$ topology is induced by
the matter of a constant energy density
with  $p=-\rho/3 =  \mp 1/(8\pi R_0^2)$
which gives the metric solution
\begin{align}\label{metric0}
ds^2 = - dt^2 +\frac{dr^2}{1 \mp r^2/R_0^2} +r^2d\Omega_2^2.
\end{align}
When the electric field is introduced in addition to this type of matter,
no consistent static solution to the Einstein's equation exists.
Therefore, the spatial dependence of the fluid matter 
was introduced as $p(r) = -\rho(r)/3$
to investigate fluid with the electric field~\cite{Cho:2017vhl}.
(The fluid-only case was investigated in Ref.~\cite{Cho:2016kpf}.)
In the region where the fluid density approaches a constant,
the geometry unveils the spatial topology. 
There are various types of solutions including black holes.

In this work, we will focus on the stability issue of the black-hole solutions
and  the static solutions without horizons.
The main purpose of this work is to perform {\it numerical} calculations
in order to verify the existence of the unstable eigen modes of the perturbations.
The stability of the solutions was studied analytically in Refs.~\cite{Cho:2016kpf,Cho:2017vhl},
and the result showed that all the solutions were unstable.
The main cause of the instability was supposed that
the fluid  evolves to drive the spacetime 
in a Friedmann-type expansion.
This instability does not necessarily mean that the black hole collapses.
The black hole may retain its stable structure 
while the instability affects the background expansion.

In order to investigate the instability in this work, 
we shall solve the perturbation equation numerically 
and search unstable modes. 
We study four cases, 
the fluid black holes with and without electric field
in $S_3$ and $H_3$.
We do the same for the static solutions without horizons.
For the $S_3$ cases, when the space is compact,
we observe the discrete spectrum of eigenvalues of unstable modes.
For the $H_3$ cases, we observe the plane-wave type of unstable modes 
with the continuous spectrum of eigenvalues.

In Sec. II, we summarize the solutions investigated in Refs.~\cite{ Cho:2017vhl,Cho:2016kpf}.
In Sec. III, we classify the solutions for numerical calculations,
discuss the boundary conditions and the numerical ranges for each class,
and present the numerical results. 
In Sec. IV, we conclude with remarking the importance of the results in a relation to future investigations.

\section{Gravitating solutions}
\label{sec-sol}
In this section, we review the solutions obtained in Ref.~\cite{Cho:2017vhl}
for the static fluid of $p(r)=-\rho(r)/3$ in the presence of the electric field.
If one turns off the electric field, 
one gets the fluid-only solutions in Ref.~\cite{Cho:2016kpf}.

The metric ansatz for spherical symmetry is given by
\begin{align}\label{metric2}
ds^2 = -f(\chi)dt^2 +g(\chi)d\chi^2 + R_0^2b^2(\chi)d\Omega_2^2,
\end{align}
where $R_0$ is a dimensionful constant,
and $b(\chi)= \sin\chi$ for $S_3$ and  $b(\chi)= \sinh\chi$ for $H_3$.
In this paper, we use the radial coordinate $\chi= b^{-1}(r/R_0)$ 
instead of $r$ because it  is more convenient to see the spatial topology
and better for numerical calculations.
The energy-momentum tensor for the fluid is given by
\begin{align}\label{emF}
T^\mu_\nu = {\rm diag}[-\rho(\chi), p(\chi),p(\chi),p(\chi)],
\end{align}
and the equation of state is
\begin{align}\label{eos}
p(\chi)=-\frac{1}{3}\rho(\chi) .
\end{align}
Since we consider the static electric field $E(\chi)$, 
the field-strength tensor ${\cal F}_{\mu\nu}$ has only the nonvanishing components,
${\cal F}_{01} = -{\cal F}_{10} = E(\chi)$. 
The energy-momentum tensor for the electric field is then given by
\begin{align}\label{emE}
{\cal T}^\mu_\nu
= {\cal F}^{\mu\alpha}{\cal F}_{\nu\alpha}
-\frac{1}{4}\delta^\mu_\nu {\cal F}_{\alpha\beta}{\cal F}^{\alpha\beta}
= \frac{E^2(\chi)}{2f(\chi)g(\chi)} {\rm diag}(-1,-1,1,1).
\end{align}
The components of the Einstein's equation are given by
\begin{align}
G^0_0 &= -\frac1{R_0^2b^2} + 2\frac{b''}{b g} +\frac{b'^2}{b^2 g} - \frac{b'g'}{b g^2}
= - 8\pi \left( \rho +\frac{E^2}{2fg} \right) ,\label{G00} \\
G^1_1 &= -\frac1{R_0^2b^2} + \frac{b'^2}{b^2 g}+ \frac{b'f'}{bfg}
= 8\pi \left( p -\frac{E^2}{2fg} \right) , \label{G11} \\
G^2_2 &= G^3_3 = \frac{b'f'}{2bfg} - \frac{f'^2}{4f^2 g} - \frac{b'g'}{2bg^2} -
	\frac{f' g'}{4f g^2} + \frac{f''}{2fg} + \frac{b''}{bg}
= 8\pi \left( p +\frac{E^2}{2fg} \right) , \label{G22}
\end{align}
where the prime denotes the derivative with respect to $\chi$. 
The energy-momentum tensors
for the fluid and the electric field are conserved individually,
$\nabla_\mu T^{\mu\nu} = 0$ and $\nabla_\mu {\cal T}^{\mu\nu} = 0$,
which provide the matter-field equations,
\begin{align}\label{TEeqn}
\rho' -\frac{f'}{f}\rho =0,
\qquad
 \frac{E'}{E} -\frac{f'}{2f} -\frac{g'}{2g} +2\frac{b'}{b}  =0.
\end{align}
From these, the matter fields are solved
in terms of the gravitational field,
\begin{align}\label{solrhoE}
\rho(\chi) = {\rm constant} \times f(\chi),
\qquad
E(\chi) = {\rm constant} \times \frac{\sqrt{f(\chi)g(\chi)}}{b^2(\chi)}.
\end{align}
With the relations in Eq.~\eqref{solrhoE},
the solutions of the Einstein's equation are presented in Tab. I.
The integration constant $K$ is related to the mass,
and $R_0$ to the curvature ~\cite{Cho:2016kpf}.
The electric field is given by
\begin{align}\label{Ftchi}
E(\chi) = \frac{Q}{|8\pi\rho_c|^{1/2}R_0^2b^2(\chi)}.
\end{align}
When $Q=0$, the solutions reduce to those for the fluid-only case in Ref.~\cite{Cho:2016kpf}.

\begin{table}
\begin{tabular}{|c||c|c|c|}
  \hline
Class & $\rho(\chi)$ & $f(\chi)$ & $g(\chi)$ \\ \hline\hline
$S_3$ 
                & $\qquad \frac{3}{8\pi R_0^2} \left[ 1- K \cot\chi -\frac{Q^2}{6R_0^2} (1-\cot^2\chi) \right] \qquad$
                & $\qquad \frac{\rho(\chi)}{\rho_c}, \quad (\rho_c>0) \qquad$
                & $\qquad \frac{3}{8\pi \rho(\chi)} \qquad$ \\ \hline
$H_3$
                & $-\frac{3}{8\pi R_0^2} \left[ 1 \mp K \coth\chi +\frac{Q^2}{6R_0^2} (1+\coth^2\chi) \right]$
                & $\frac{\rho(\chi)}{\rho_c}, \quad (\rho_c<0)$
                & $-\frac{3}{8\pi \rho(\chi)}$ \\
  \hline
\end{tabular}
\caption{Classification of solutions.
The signature of $\rho_c$ is chosen so that $f(\chi)g(\chi)>0$.
}
\end{table}

\subsection{$S_3$ solution}
The metric for $S_3$ is given by
\begin{equation}\label{metricS3I}
ds^2 = -\frac{3}{8\pi R_0^2\rho_c} \left[ 1- K \cot\chi
-\frac{Q^2}{6R_0^2} (1-\cot^2\chi) \right] dt^2
+\frac{R_0^2}{1- K \cot\chi
-(Q^2/6R_0^2) (1-\cot^2\chi)} d\chi^2
+R_0^2\sin^2\chi d\Omega_2^2.
\end{equation}

\subsubsection{Charged case}
This is the case of $Q \neq 0$.
When $J_1 \equiv 9K^2R_0^4 -6Q^2R_0^2+Q^4>0$,
the solution describes a black hole of Reissner-Nortstr\"om (RN) type.
The $S_3$ topology appears about $\chi =\pi/2$ with the charge correction.
There exist two horizons at 
$\chi_\pm = \cot^{-1} [ (3KR_0^2 \mp \sqrt{J_1})/Q^2]$,
between which the spacetime is nonstatic.
There exist two singularities at $\chi=0$ and $\chi=\pi$.
The former is not accessible by timelike observers as in the RN black hole.
The latter is naked but is not accessible either by the timelike observers
as studied in terms of geodesics in Ref.~\cite{Cho:2017vhl}.

When $J_1 <0$, the solution becomes static in all space.
The horizons disappear and the singularities become naked.

\subsubsection{Fluid-only case}
This is the case of $Q = 0$.
In this case, there exists only a black-hole solution.
The horizon is located at $\chi_h = \cot^{-1}(1/K) <\pi/2$,
inside which the spacetime is nonstatic.
The story of singularities is the same as the charged case.

\subsection{$H_3$ solution}
The metric for $H_3$ is given by
\begin{equation}\label{metricH3}
ds^2 = -\frac{3}{8\pi R_0^2(-\rho_c)} \left[ 1- K \coth\chi
+\frac{Q^2}{6R_0^2} (1+\coth^2\chi)  \right] dt^2
+\frac{R_0^2}{1 - K \coth\chi +(Q^2/6R_0^2) (1+\coth^2\chi)} d\chi^2
+R_0^2\sinh^2\chi d\Omega_2^2.
\end{equation}

\subsubsection{Charged case}
When $J_2 \equiv9K^2R_0^4 -6Q^2R_0^2 -Q^4>0$
and $3(K-1)R_0^2 < Q^2 < 3KR_0^2$,
the solution describes RN-type black hole.
There are two horizons at
$\chi_\pm = \coth^{-1} [ (3KR_0^2 \mp \sqrt{J_2})/Q^2]$.
In the region between the horizons,
the spacetime is nonstatic.
The $H_3$ topology appears as $\chi \to \infty$ with the mass and charge corrections.
There is a curvature singularity at $\chi=0$.

When $J_2 <0$, or $J_2>0$ and $Q^2 > 3KR_0^2$, or $K<0$,
the horizons disappear and the solution becomes static.
The singularity at the center is naked.

\subsubsection{Fluid-only case}
If $K>0$, the solution describes a black hole.
The horizon is located at $\chi_h = \coth^{-1}(1/K)$,
inside which the spacetime is nonstatic.
The story of singularities is the same as the charged case. 

If $K<0$, the horizon disappears and the spacetime becomes static.
The singularity at the center becomes naked.

\section{Numerical Study of Stability}
In this section, we study numerically the stability of the black-hole  solutions 
and the static solutions without horizons.
We adopt the same linear spherical scalar perturbations in Ref.~\cite{Cho:2017vhl}.
With the metric ansatz,
\begin{align}
ds^2 = -f(t,\chi)dt^2 + g(t,\chi) d\chi^2 + R_0^2 b^2(\chi) d\Omega_2^2,
\end{align}
we introduce the radial metric perturbations as
\begin{align}
f(t,\chi) &= f_0(\chi) + \epsilon f_1(t,\chi), \label{p1}\\
g(t,\chi) &= R_0^2 \big[ g_0(\chi) + \epsilon g_1(t,\chi) \big], \label{p2}
\end{align}
where $\epsilon$ is a small parameter.
The subscript $0$ stands for the background solutions
summarized in Sec. II.
Let us define $F(\chi) \equiv 1/g_0(\chi) = 8\pi R_0^2\rho_0(\chi)/3s$ for convenience,
where $s=+1(-1)$ for $S_3(H_3)$, and 
$\rho_0(\chi)$ is the background solution in Tab. I.
The energy-momentum tensor for fluid is given by
\begin{align}
T^{\mu\nu} = (\rho +p)u^\mu u^\nu + pg^{\mu\nu},
\end{align}
where the velocity four-vector is
\begin{align}
u^\mu = \big[ u^{t}(t,\chi),u^{\chi}(t,\chi),0,0 \big].
\end{align}
For fluid with $p=-\rho/3$,
the perturbations for the energy density and the four-velocity are introduced by
\begin{align}
\rho(t,\chi) &= \rho_0(\chi) +\epsilon \rho_1(t,\chi), \label{p3}\\
u^{t}(t,\chi) &= u_0^{t}(\chi) +\epsilon u_1^{t}(t,\chi), \label{p4}\\
u^{\chi}(t,\chi) &= u_0^{\chi}(\chi) +\epsilon u_1^{\chi}(t,\chi). \label{p5}
\end{align}
We have $u_0^{\chi}(\chi)=0$ for the comoving background fluid.
The normalization condition $u^\mu u_\mu=-1$ gives
$u_0^{t}(\chi)=1/\sqrt{f_0(\chi)}$ and
$u_1^{t}(t,\chi)=-f_1u_0^{t}/(2f_0) = -f_1/(2f_0^{3/2})$.

For the electric field, we consider the perturbation
only along the radial direction,
by which there is no magnetic field induced by the perturbation,
\begin{align}\label{p6}
{\cal F}'_{t\chi} = -{\cal F}'_{\chi t} = E(t,\chi) = E_0(\chi) + \epsilon E_1(t,\chi),
\end{align}
where $E_0(\chi)$ is given in Eq.~\eqref{Ftchi}.

We apply the perturbations, \eqref{p1}, \eqref{p2}, and \eqref{p3}-\eqref{p6},
on the field equations.
In the first order of $\epsilon$,
the $(0,1)$ component of the Einstein's equation gives
\begin{align}
u_1^{\chi}(t,\chi) = -\sqrt{\frac{2\pi R_0^2\rho_c}{3}} \frac{\dot{g_1}b'F}{b\sqrt{F}}.
\end{align}
We note that the perturbations of the four-vector, $u_1^{t}$ and $u_1^{\chi}$ in Eqs.~\eqref{p4} and \eqref{p5},
are expressed completely by the metric perturbations and the background functions.
While we have four perturbation functions,$f_1$, $g_1$, $\rho_1$ and $E_1$,
there are seven equations; 
three from the diagonal components of the Einstein's equation, 
two from $\nabla_\mu T^{\mu\nu} =0$ ,
and two from $\nabla_\mu {\cal T}^{\mu\nu} =0$.
Therefore, three of them are redundant.
We synchronize the metric perturbations as
\begin{align}\label{f1g1}
f_1(t,\chi) = e^{i\omega t} \psi(\chi),\qquad
g_1(t,\chi) = e^{i\omega t} \varphi(\chi),
\end{align}
where the constant $w$ is the frequency of the perturbation.
We can decouple the equation for $\varphi(\chi)$ as
\begin{align}\label{PE1}
-F^2\varphi''
-\left[ 3FF' + F^2 \left( 3\frac{b''}{b'} +s\frac{b}{b'} \right) \right] \varphi'
+\left[ \frac{\omega^2}{\sigma} -2FF''
-FF'\left( 4\frac{b''}{b'}-\frac{b'}{b} -s\frac{b}{b'} \right)
-2F^2 \left( \frac{b'''}{b'} -\frac{b'^2}{b^2} +s \frac{bb''}{b'^2} -s \right)
\right] \varphi =0,
\end{align}
where $\sigma \equiv 1/(8\pi R_0^4\rho_c s) = 1/(8\pi R_0^4|\rho_c|) >0$.
The coefficients of this equation depend only on the background functions $F(\chi)$ and $b(\chi)$. As only one perturbation function is relevant, we could ignore the equation for $\psi(\chi)$.

Performing transformations for the radial coordinate and the amplitude function as
\begin{align}\label{z}
z =\pm \int^\chi \frac{d\chi}{\sqrt{2}F(\chi)} +z_0, \qquad
\Psi(z) = N F(\chi)b'(\chi) \varphi(\chi) ,
\end{align}
where $N$ is a normalization constant,
Eq.~\eqref{PE1} can be cast in the nonrelativistic Schr\"odinger-type,
\begin{align}\label{PE2}
\left[ -\frac{1}{2}\frac{d^2}{dz^2} 
+U(z) \right]\Psi(z)= -\frac{\omega^2}{\sigma}\Psi(z)
=-8\pi R_0^4|\rho_c| \omega^2 \Psi(z) \equiv \Omega \Psi(z),
\end{align}
where the potential is given by~\footnote{Here, we correct the typos in Ref.~\cite{Cho:2017vhl}.}
\begin{align}\label{U}
U[z(\chi)] = F^2 \left[ -\frac{F''}{F} +\left( \frac{F'}{F} \right)^2
+\frac{F'}{F} \left(-2 \frac{b''}{b'} +\frac{b'}{b}  \right)
+2 \left( \frac{b''}{b'} \right)^2  +2\left( \frac{b'}{b}  \right)^2 +3s
\right].
\end{align}
Here, the prime denotes the derivative with respect to $\chi$, 
and the relation $sb=-b''$ is used to arrange terms in the ordered way. 
The relation between the amplitude functions is $\Psi(z)=z\Phi(z)$; 
$\Phi(z)$ used in Ref.~\cite{Cho:2017vhl} 
is the wave function of the three dimensional form of the Schr\"odinger equation, 
while $\Psi(z)$ is that of the one dimensional form 
of which the squared value represents the distribution  function.
Although we solve Eq.~\eqref{PE1}, we interpret the solution as an eigenvalue problem in quantum mechanics
using  Eqs.~\eqref{PE2} and \eqref{U}.
The potential $U(\chi)$ is plotted in Figs. 1 and 2.
Regardless of the shape of the potential, 
there always exists a positive eigenvalue $\Omega$, i.e., $\omega^2 <0$.
This indicates that the system is unconditionally unstable. 

The cause of instability is two folds.
First, the perturbation in fluid can induce the Friedmann expansion on the spacetime.
This does not necessarily destroy the black-hole structure.
Second, the perturbation in the electric field may cause 
the destruction of the black-hole structure
known as the instability of Cauchy (inner) horizon \cite{Gursel:1979zza}.

\clearpage
\begin{figure*}[btph]
\begin{center}
\includegraphics[width=0.3\textwidth]{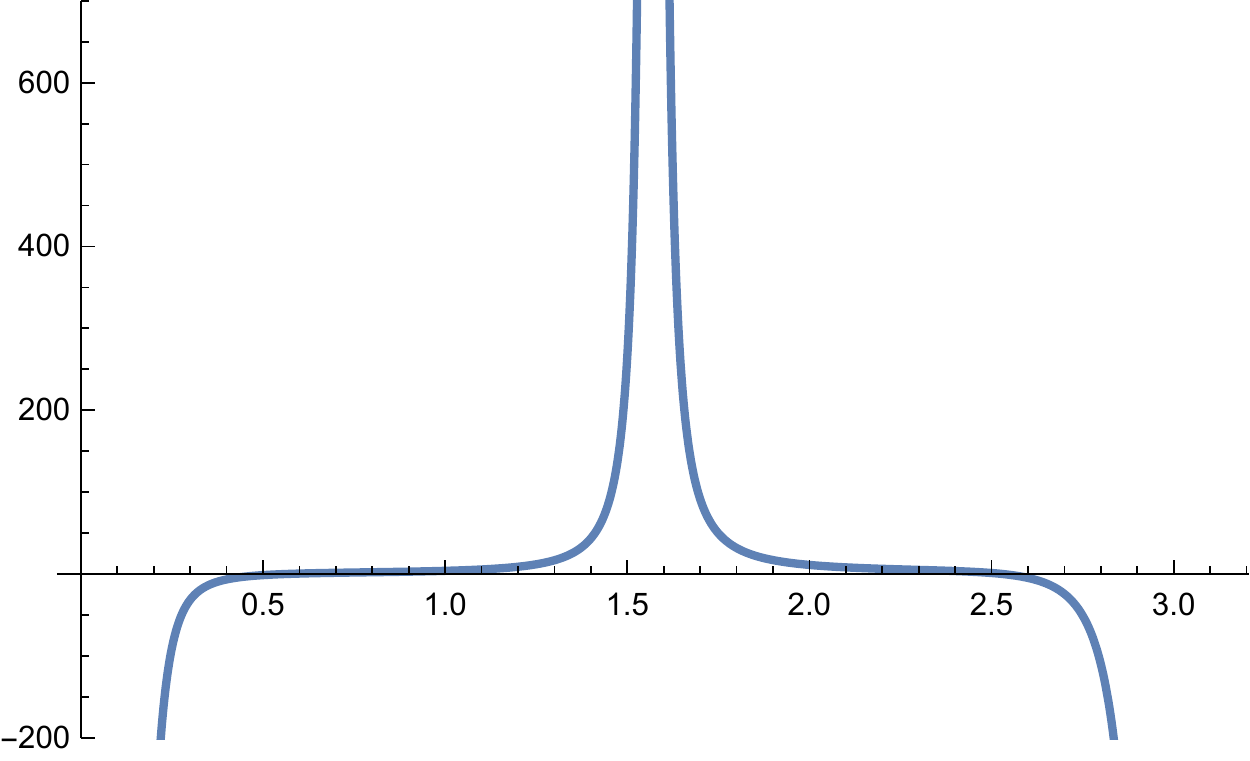}
\includegraphics[width=0.3\textwidth]{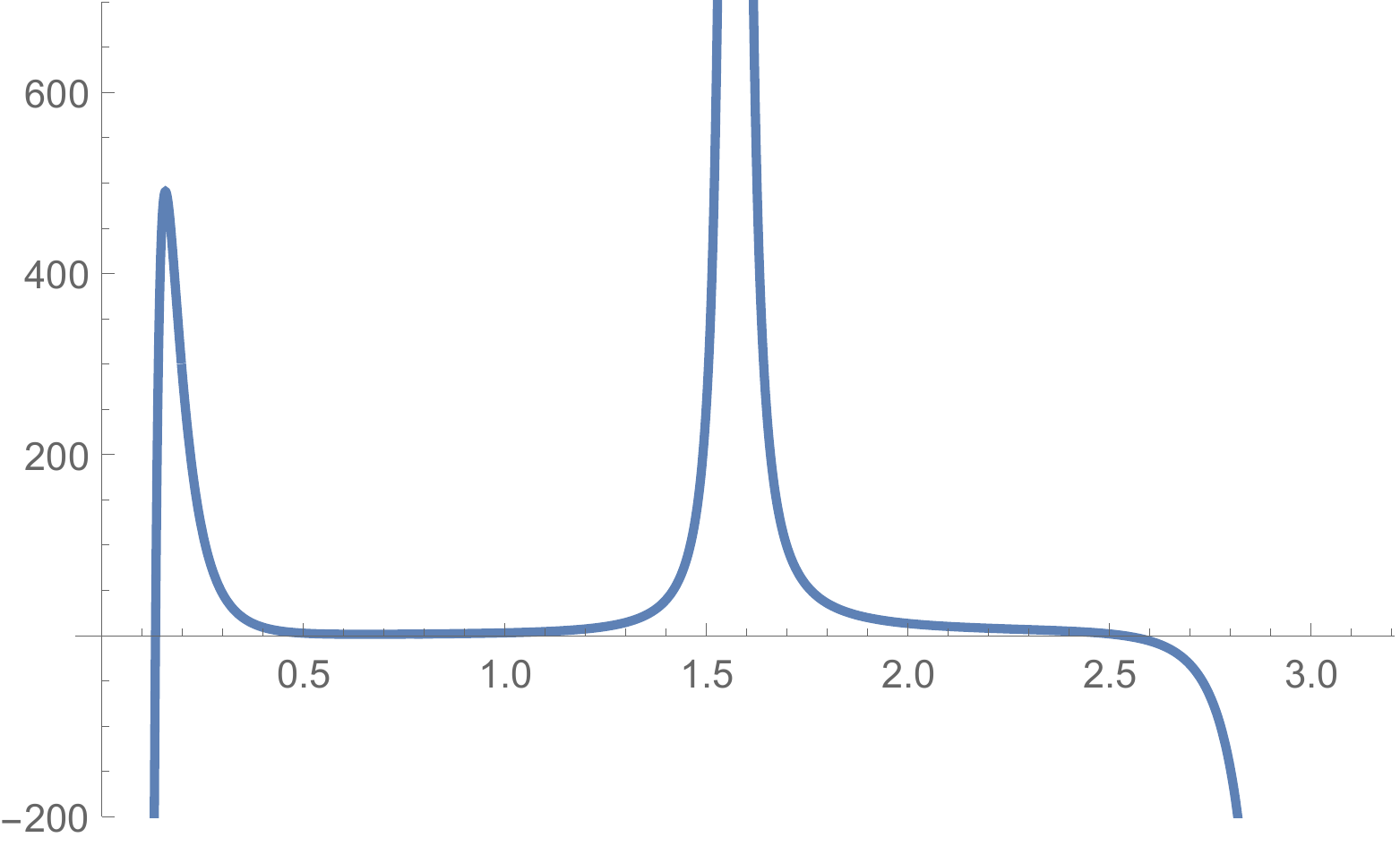}
\includegraphics[width=0.3\textwidth]{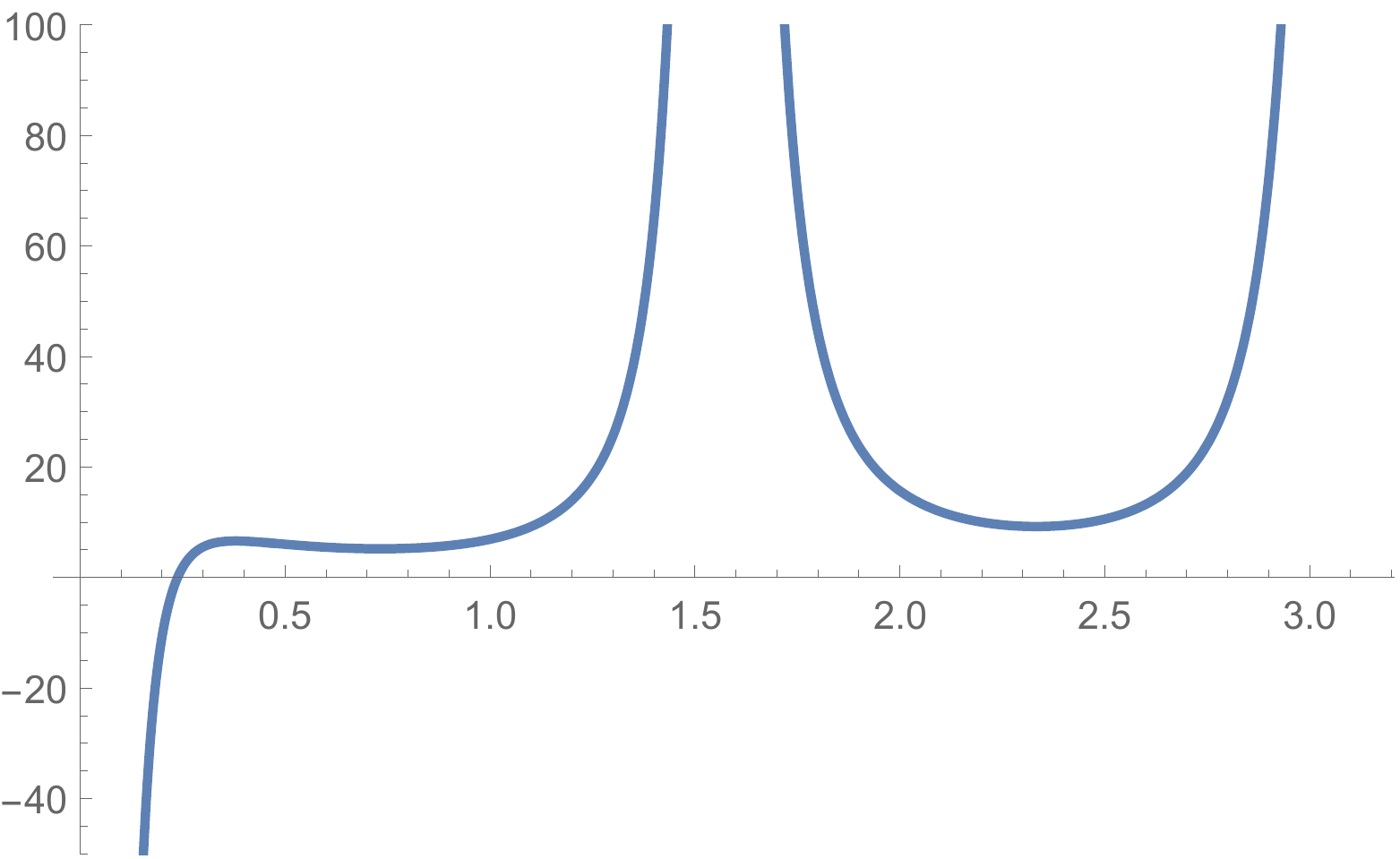}\\
(a) SST1 \hspace{1.7in} (b) SBH1 \hspace{1.7in} (c) SBH2\\
\end{center}
\caption{Plot of potential $U(\chi)$ for $S_3$.
The potential blows up at $\chi = 0, \pi/2, \pi$.
(a) Charged static case for $K=5/9$, $Q=1$, $R_0=1$.
There is no horizon. 
(b) Charged black-hole case for $K=1$, $Q=1$, $R_0=1$.
There are two horizons.
(c) Fluid-only black-hole case for $K=0.5$.
There is one horizon.
The numerical calculation will be performed in the range,
(a) $[0,\pi/2)$, (b) $[0,\chi_-)$ and $(\chi_+,\pi/2]$, (c) $(\pi/2,\pi]$.
}
\end{figure*}
\vspace{12pt}
\begin{figure*}[btph]
\begin{center}
\includegraphics[width=0.3\textwidth]{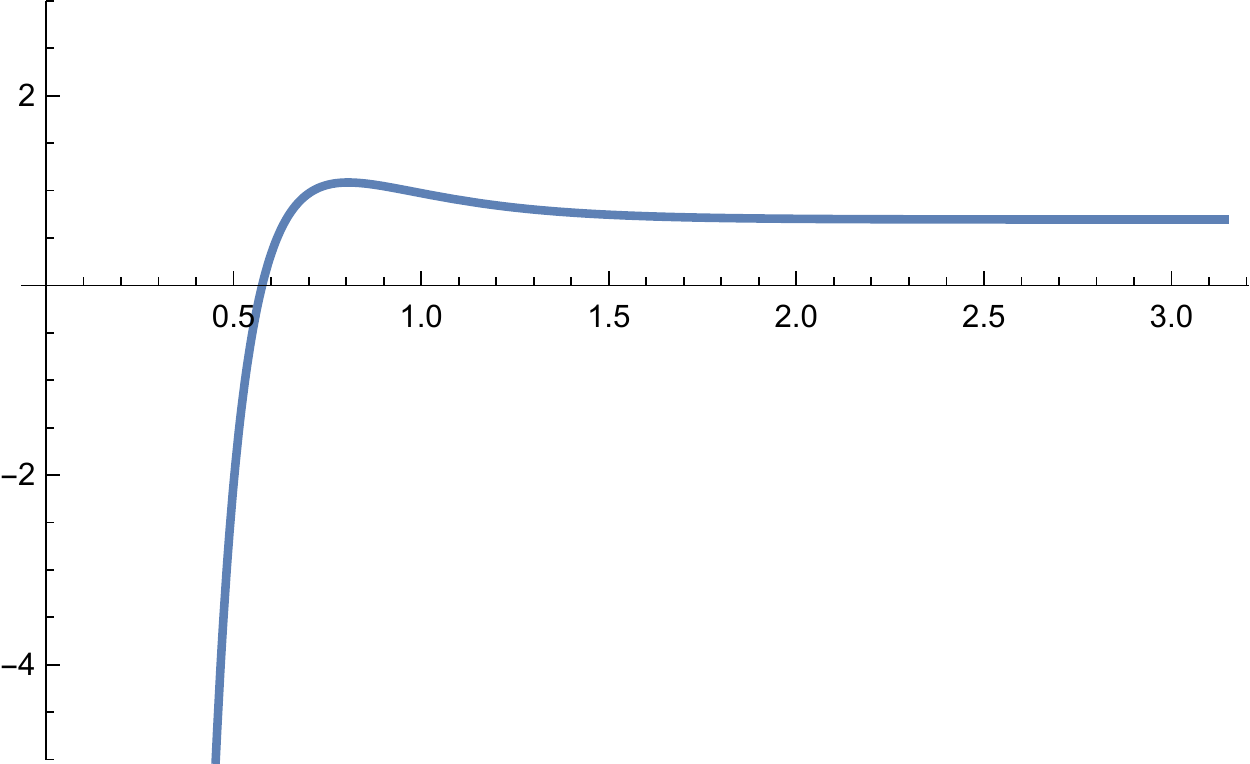}
\includegraphics[width=0.3\textwidth]{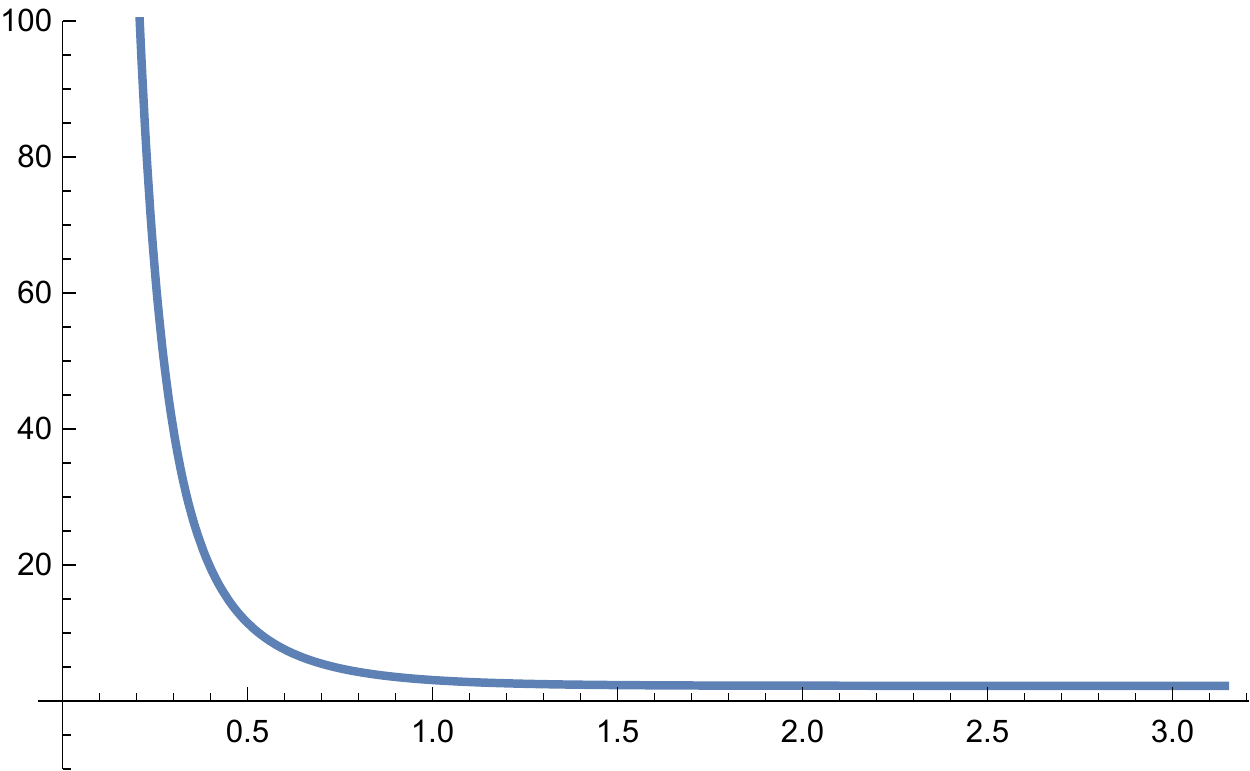}\\
(a)  HST1 \hspace{1.7in} (b) HST2 \\
\includegraphics[width=0.3\textwidth]{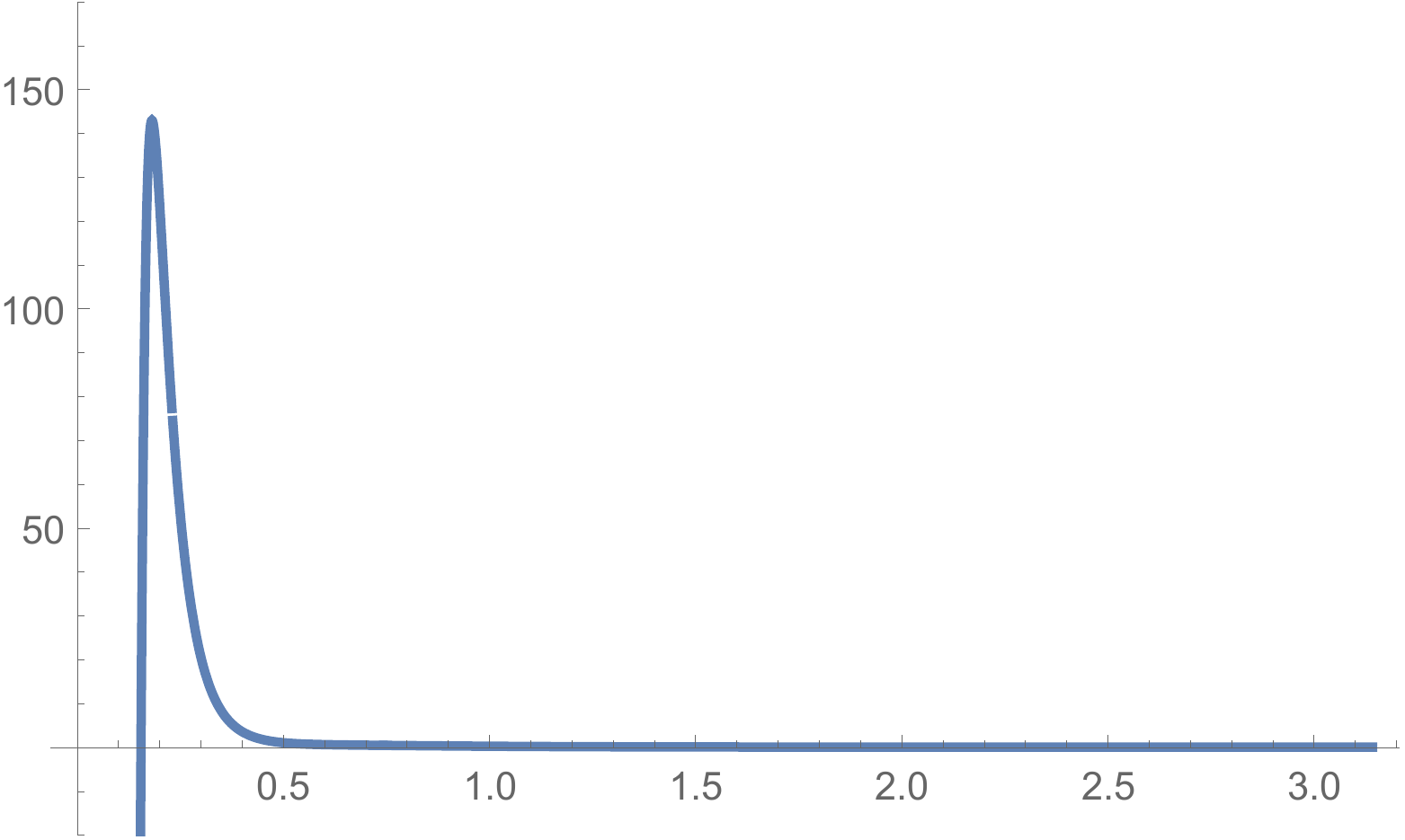}
\includegraphics[width=0.3\textwidth]{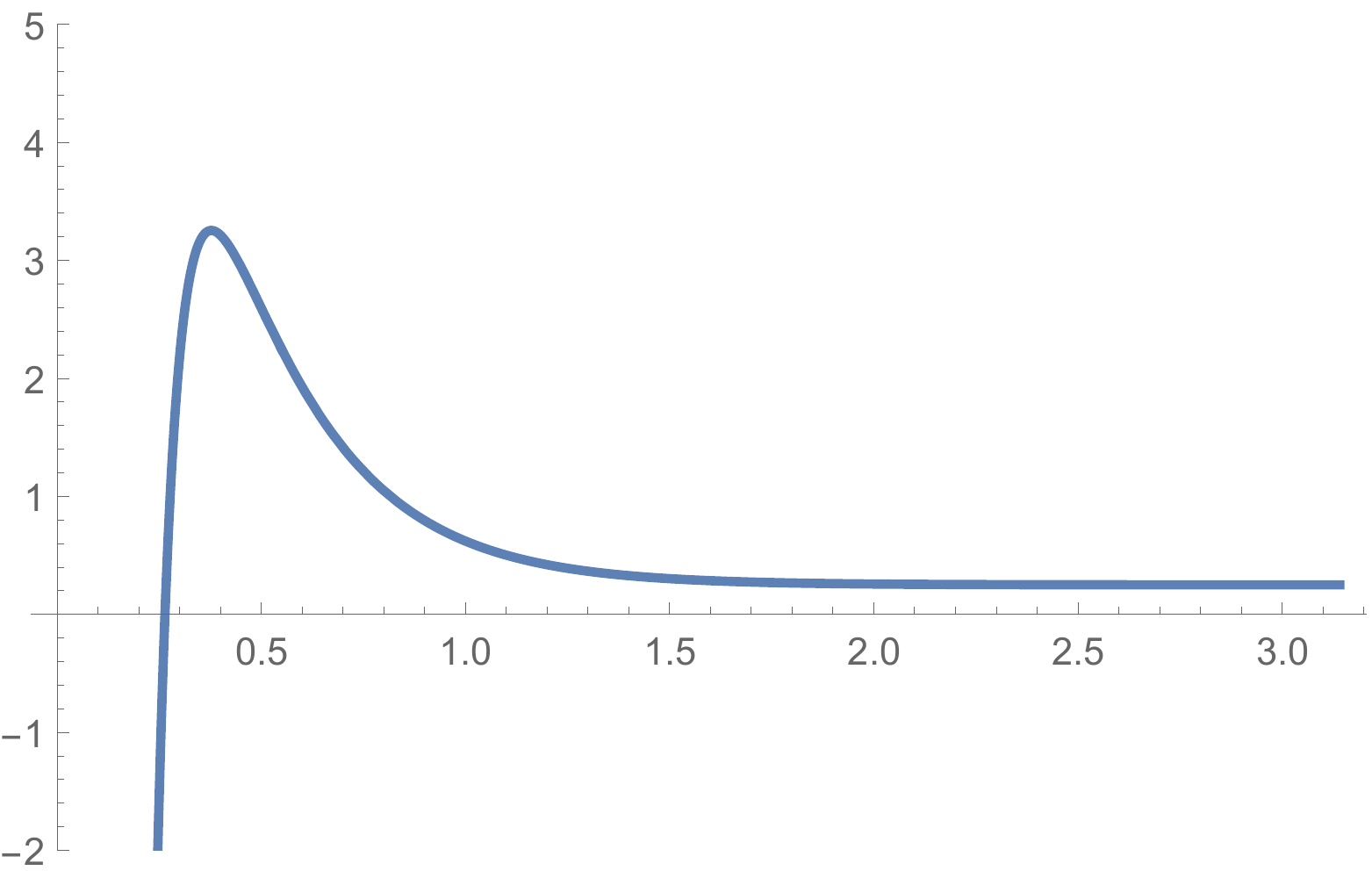}\\
(c)  HBH1 \hspace{1.7in} (d)  HBH2 \\
\end{center}
\caption{Plot of potential $U(\chi)$ for $H_3$.
The potential blows up at $\chi = 0$.
(a) Charged static case for $K=0.5$, $Q=1$, $R_0=1$.
There is no horizon.
(b) Fluid-only static case for $K=0.5$.
There is no horizon.
(c) Charged black-hole case for $K=1$, $Q=1$, $R_0=1$.
There are two horizons.
(d) Fluid-only black-hole case for $K=-0.5$.
There is one horizon.
The numerical calculation will be performed in the range,
(a) $[0,\chi_b]$, (b) $[0,\chi_b]$, (c) $[0,\chi_-)$ and $(\chi_+,\chi_b]$,  (d) $(\chi_h,\chi_b]$.
}
\end{figure*}

\clearpage
\begin{table}
\begin{tabular}{|c||c|c||c|c|}
  \hline
Class & static (charged)  & static (fluid-only)  & black hole (charged) & black hole (fluid-only) \\ \hline\hline
 $S_3$
                & SST1: $[0,\pi/2)$
                & SST2: $\text{not available}$
                & SBH1-I: $[0,\chi_-)$, SBH1-II: $(\chi_+,\pi/2]$ 
                & SBH2: $(\pi/2,\pi]$ \\ \hline
$H_3$
                & HST1: $[0,\chi_b] $
                & HST2: $[0,\chi_b] $
                & HBH1-I: $[0,\chi_-)$, HBH1-II: $(\chi_+,\chi_b]$ 
                & HBH2: $ (\chi_h,\chi_b]$ \\
  \hline
\end{tabular}
\caption{Range for numerical calculations ($\chi_b$: an arbitrarily large value, 
$\chi_-$: the inner horizon of the charge black hole,
$\chi_+$: the outer horizon of the charge black hole,
$\chi_h$: the horizon of the black hole).
In the range, the spacetime is static, $f(\chi)>0$ and $g(\chi)>0$.
}
\end{table}

\subsection{Numerical ranges and boundary conditions}
In this work, we solve Eq.~\eqref{PE1} numerically in order to obtain
the eigenvalues $\Omega =-w^2/\sigma$ 
and the corresponding eigenfunctions $\varphi(\chi)$, thereby $\Psi[z(\chi)]$.
Let us discuss the range of numerical calculations and boundary conditions.
Since we introduced the perturbations as Eq.~\eqref{f1g1}
for which $t$ and $\chi$ are the temporal and the spatial coordinates respectively,
we perform numerical calculations in the {\it static} region of the spacetime
where  $f(\chi)>0$ and $g(\chi)>0$.

We can classify the solutions in eight  for numerical calculations as in Tab. II.
For $S_3$ and $H_3$,
we have the {\it charged} and the {\it fluid-only} cases,
and each case has the {\it static} and the {\it black-hole} solutions.
Only $S_3$ fluid-only static solution does not exist,
so we have only seven classes in total.

Let us denote the left and the right boundaries as $\chi_a$ and $\chi_b$, respectively.
For the {\it static} solution (SST1, HST1, HST2),
the spacetime is static in the entire region.
For the {\it charged} black-hole solution (SBH1, HBH1),
the spacetime is static inside the inner horizon ($\chi<\chi_-$)
and outside the outer horizon ($\chi>\chi_+$).
For the {\it uncharged} black-hole solution  (SBH2, HBH2),
the spacetime is static outside the horizon ($\chi>\chi_h$).

In determining the boundares, $\chi_a$ and $\chi_b$,
we also consider the shape of the potential $U$ in Figs. 1 and 2.
The ranges are summarized in Tab. II.
Below, we describe more in detail by class how to determine the ranges.
We also present boundary conditions $\varphi_a \equiv \varphi(\chi_a)$
and $\varphi_b \equiv \varphi(\chi_b)$.

\vspace{12pt}
(SST1) $S_3$ static (charged):
The spacetime is static in the entire region of sphere, $\chi=[0,\pi]$.
For the $S_3$ class, the potential $U$ diverges at $\chi=\pi/2$.
When it is charged, 
if we perform the series expansion for the coefficients of Eq.~\eqref{PE1},
and take the most dominant terms,
the solution about $\chi=\pi/2$ is given by
\begin{align}\label{BCpi2}
\varphi(\chi \approx \pi/2) &=
c_1 \left[ {\cal C}_{11}(K,Q,R_0)
\left( \chi - \frac{\pi}{2} \right)^{-2}
+ {\cal C}_{12}(K,Q,R_0)
\left( \chi - \frac{\pi}{2} \right)^{-1} + \cdots
\right] \nonumber\\
&+ c_2 \left[ {\cal C}_{21}(K,Q,R_0)
\left( \chi - \frac{\pi}{2} \right) 
+{\cal C}_{22}(K,Q,R_0)
\left( \chi - \frac{\pi}{2} \right)^2+ \cdots
\right],
\end{align}
where ${\cal C}_{ij}$'s are constants depending on the parameters $K$, $Q$, and $R_0$
($\omega$ dependence of ${\cal C}_{ij}$ appears in the next order).
The solution possesses divergent terms at $\chi=\pi/2$.
However, one can see from Eq.~\eqref{z} 
that the wave function cannot be normalized with the divergent term.
Therefore, we set $c_1=0$, and we have $\varphi(\pi/2)=0$.
We take the numerical range for SST1 as $\chi = [0,\pi/2)$.~\footnote{We 
perform the numerical calculation out to $\chi= \pi/2 -\epsilon$,
where $\epsilon$ is a small number.
If we take the range as $\chi = [0,\pi/2]$,
with the boundary conditions $\varphi(0)=\varphi(\pi/2)=0$,
the numerical result gives only a trivial solution, $\varphi(\chi)=0$.}
For the charged cases (SST1, SBH1-I, HST1, HBH1-I), 
one can impose the boundary conditions as
\begin{align}\label{BC}
\varphi(\chi_a=0) = 0, \qquad
\varphi(\chi_b) = \varphi_b,
\end{align}
where $ \varphi_b$ is a nonzero constant
which is free to rescale
since Eq.~\eqref{PE1} is linear in $\varphi(\chi)$. 
The boundary condition $\varphi(\chi_a=0) = 0$ is guaranteed as following.
Performing the series expansion for the coefficients of Eq.~\eqref{PE1} about $\chi=0$, 
and taking the most dominant terms,
we get the solution,
\begin{align}
\varphi(\chi) \approx \chi^3 \left( c_3 + c_4 e^{18KR_0^2\chi/Q^2} \right),
\end{align}
which gives  $\varphi(\chi\to 0) \to 0$.


\vspace{12pt}
(SST2) $S_3$ static (fluid-only): The solution does not exist.

\vspace{12pt}
(SBH1) $S_3$ black hole (charged):
We perform numerical calculations at two static regions;
inside the inner horizon,  $\chi = [0,\chi_-) $,
and outside the outer horizon, $\chi = (\chi_+, \pi/2]$.
Two horizons are always located inside the equator, $\chi_\pm< \pi/2$,
at which the potential $U$ diverges.
The boundary conditions in Eq.~\eqref{BC} are applied for inside the inner horizon (SBH1-I). 
For outside the outer horizon (SBH1-II), we impose the boundary conditions as
\begin{align}\label{BCout}
\varphi(\chi_a=\chi_+ +\epsilon) = \varphi_a, \qquad
\varphi(\chi_b=\pi/2) = 0,
\end{align}

\vspace{12pt}
(SBH2) $S_3$ black hole (fluid-only):
The spacetime is static outside the horizon
which is located always inside the equator, $\chi_h < \pi/2$.
Since the potential $U$ diverges at the equator and at the south pole ($\chi=\pi$),
we take the numerical range as $\chi = (\pi/2, \pi]$.~\footnote{
We may take the numerical range as $(\chi_h,\pi/2]$ with boundary conditions, 
$\varphi(\chi_a=\chi_h +\epsilon) = \varphi_a$ and $\varphi(\chi_b=\pi/2) = 0$.
For this case, the situation will be very similar to that of SBH1-II,
and the result as well.}
If we perform the series expansion for the coefficients of Eq.~\eqref{PE1},
and take the most dominant terms,
the solutions at the boundaries become
\begin{align}\label{SBH2}
\varphi(\chi \approx \pi/2) &=
c_5 \left[ {\cal C}_{51}(K)
\left( \chi - \frac{\pi}{2} \right)^{-2}
+ {\cal C}_{52}(K)
\left( \chi - \frac{\pi}{2} \right)^{-1} + \cdots
\right] 
+ c_6 \left[ {\cal C}_{61}(K)
\left( \chi - \frac{\pi}{2} \right) 
+{\cal C}_{62}(K)
\left( \chi - \frac{\pi}{2} \right)^2+ \cdots
\right],\nonumber\\
\varphi(\chi \approx \pi) 
&=c_7 \left[ {\cal C}_{71}(K)
\left( \chi -\pi \right)
+ {\cal C}_{72}(K)
\left( \chi - \pi \right)^2 + \cdots
\right] 
+ c_8 \left[ {\cal C}_{81}(K)
\left( \chi - \pi \right)^3 
+{\cal C}_{82}(K)
\left( \chi -\pi\right)^4+ \cdots
\right].
\end{align}
Similarly to Eq.~\eqref{BCpi2}, 
the divergent term should be removed by setting $c_5=0$
on the normalization purpose.
Therefore, we take the boundary conditions as
$\varphi(\pi/2 + \epsilon) = \varphi_a$ and $\varphi(\pi) = 0$.

\vspace{12pt}
(HST1) $H_3$ static (charged):
Since the spacetime is static in the entire region, $\chi=[0,\infty)$,
we take the outer boundary $\chi_b$ at a large value 
where the potential $U$ approaches a constant.
The boundary conditions in Eq.~\eqref{BC} are applied for this class.

\vspace{12pt}
(HST2) $H_3$ static (fluid-only):
The situation is the same with HST1.
The boundary conditions in Eq.~\eqref{BC} are applied also for this class.
The boundary condition $\varphi(\chi_a=0) = 0$ is guaranteed as following.
Performing the series expansion for the coefficients of Eq.~\eqref{PE1} about $\chi=0$, 
we get the solution,
\begin{align}
\varphi(\chi) \approx c_9(K\chi + 2\chi^2) +c_{10}\chi^3,
\end{align}
which gives  $\varphi(\chi\to 0) \to 0$.

\vspace{12pt}
(HBH1) $H_3$ black hole (charged):
Similarly to SBH1, 
we perform numerical calculations in two regions;
inside the inner horizon,  $\chi = [0,\chi_-) $ 
and outside the outer horizon, $\chi = (\chi_+, \chi_b]$, where $\chi_b$ is a large value.
The boundary conditions in Eq.~\eqref{BC} are applied for inside the inner horizon (HBH1-I). 
At large $\chi$, the solution $\varphi$ behaves as a plane wave.
Therefore, $\varphi_b$ is not fixed to a specific value
because of two arbitrary integration constants.
Near the horizon, $\varphi_a$ is not fixed either due to the same reason.
Therefore, the boundary conditions for outside the outer horizon (HBH1-II) 
are not well fixed at both boundaries.

\vspace{12pt}
(HBH2) $H_3$ black hole (fluid-only):
Since the spacetime is static outside the horizon,
we take the numerical range as
$\chi = (\chi_h,\chi_b]$, 
where the location of horizon is $\chi_h = \coth^{-1}(1/K)$
and $\chi_b$ is a large value.
We perform the series expansion for the coefficients of Eq.~\eqref{PE1}.
Similarly to HBH1-II, the boundary conditions at both boundaries are not well fixed for this case.


\subsection{Numerical results}
Imposing the boundary conditions discussed in the previous subsection,
we solve  Eq.~\eqref{PE1} numerically to obtain $\varphi(\chi)$
in the numerical ranges in Tab. II.
We replace the ordinary differential equation with second-order finite-difference equations, 
and then use the {\it relaxation method} with Newton's iteration scheme 
by solving the inverse of a banded matrix~\cite{Park:2013yka}.
We plot the results in Figs. 3 and 4.
We obtained several eigenvalues of $\Omega=-w^2>0$ (we set $\sigma=1$),
and plotted corresponding eigenfunctions $\Psi(z)$. 

\vspace{12pt}
(SST1) $S_3$ static (charged):
The numerical range $\chi=[0,\pi/2)$ corresponds to $z=[0,z_b)$
where $z_b$ is a finite value.~\footnote{From Eq.~\eqref{z}, 
we can set the value of the integration constant $z_0$ 
so that the left boundary is located at $z=0$.}
Considering the potential $U(z)$ in Fig. 5 in the finite range,
the solutions exhibit a tower of discrete eigenvalues as expected.
The ground state (in black) is obtained for $\Omega<0$
which is a {\it stable} mode,
while the rest of excited modes are unstable modes, $\Omega >0$.
We can observe that the number of  the node in the eigenfunction increases
as the eigenvalue increases.

\vspace{12pt}
(SBH1-I) $S_3$ black hole (charged) inside the inner horizon:
The numerical range $\chi=[0,\chi_-)$ corresponds to $z=[0,\infty)$ which is infinite.
As $\chi$ approaches the inner horizon, $z$ diverges.
The potential $U$ in this range possesses a bump approaching a constant value
as $z$ increases to infinity.
Since the numerical range in $z$ is infinite with this potential,
the solutions exhibit the plane-wave behaviour at large $z$.
It has a continuous spectrum of eigenvalues.
The divergence of $z(\chi)$ is very steep near the horizon,
so the oscillation of the wave function becomes very rapid 
in the very short range of $\chi$ near $\chi_-$.
This induces numerical difficulties,
so we perform the numerical calculations in the inner finite region in $z$,
$\chi=[0,\chi_b (< \chi_-)] \to z=[0,z_b]$,
where these difficulties can be avoided.

\vspace{12pt}
(SBH1-II) $S_3$ black hole (charged) outside the outer horizon:
The numerical range $\chi=(\chi_+,\pi/2]$ corresponds to $z=(-\infty,0]$ 
by adjusting the integration constant $z_0$ so that $z(\chi=\pi/2)=0$.
As $\chi$ approaches the inner horizon, $z$ diverges. 
We changed the signature of $z$
to make the range of $z$ be $[0,\infty)$ for convenience with other cases. 
The potential $U(z)$ in this range is plotted in Fig. 5.
Since the numerical range in $z$ is infinite with this potential,
the solutions exhibit the plane-wave behaviour at large $z$.
It has a continuous spectrum of eigenvalues.

\vspace{12pt}
(SBH2) $S_3$ black hole (fluid-only):
The numerical range $\chi =(\pi/2,\pi]$ corresponds to $z =(z_a,z_b]$ which is finite. We adjusted the integration constant $z_0$ so that $z(\chi=\pi)=0$ and changed the signature of $z$
to make the range of $z$ be $[0,z_b)$ for convenience with other cases.
The potential $U$ in this range is a potential well.
The solutions exhibit a tower of discrete eigenvalues which are all unstable modes.
Although we imposed the vanishing boundary condition at the right boundary, $\varphi (\chi=\pi) =0$,
the value of the wave function $\Psi$ is a nonzero value;
using the solution at the boundary in Eq.~\eqref{SBH2},
we get $\Psi$ at the right boundary from Eq.~\eqref{z},
\begin{align}\label{SBH2bc}
\Psi(\chi) = NF(\chi)b'(\chi) \varphi(\chi)
= c_7NK -c_7N(\chi-\pi) + \left( c_8-\frac{5}{6}\right) NK (\chi-\pi)^2 + \cdots
\qquad\Rightarrow\qquad
\Psi(\pi) = c_7NK.
\end{align}

\vspace{24pt}
(HST1) $H_3$ static (charged):
The numerical range  $\chi=[0,\chi_b]$ corresponds to $z =[0,z_b]$.
The situation is very similar to SBH1-I.
The potential $U$ has a  bump and approaches a constant value as $z(\chi)$ increases 
($\chi$ and $z$ are unbounded).
The solutions exhibit the plane-wave behaviour
with a continuous spectrum of eigenvalues larger than the asymptotic value of the potential $U$. 

\vspace{12pt}
(HST2) $H_3$ static (fluid-only):
The situation is the same with HST1,
except that the potential $U$ is divergent oppositely at the origin.

\vspace{12pt}
(HBH1-I) $H_3$ black hole (charged) inside the inner horizon:
The situation is the same with SBH1-I.

\vspace{12pt}
(HBH1-II) $H_3$ black hole (charged) outside the outer horizon: We perform the numerical calculation outside the horizon. The numerical range  $\chi=(\chi_+,\chi_b]$ corresponds to $z =[z_a,z_b]$. 
At the horizon, $z(\chi \to \chi_+) \to -\infty$ and the potential $U$ approaches a finite value. $\chi$ and $z$ are unbounded outwards,
but we fix the numerical boundary at large $z_b = z(\chi_b)$
where the potential approaches a constant value.
In this range, the potential $U(z)$ is a decreasing function
from a finite value as in Fig. 6.
As it was discussed in the previous section,
at both boundaries, the boundary values are not well fixed.
However, we performed the numerical calculation
with arbitrary nonzero boundary conditions.
As it was expected from the potential shape,
we could obtain the plane-wave type solutions for any eigenvalues
larger than the value of the potential $U$. 
Because of the arbitray boundary conditions, 
phase constants of different eigenvalue solutions also are arbitrary. 
Here, we translate $z$ coordinate, denoted $z^*$, to set phase constants of different eigenvalue solutions equal.

\vspace{12pt}
(HBH2) $H_3$ black hole (fluid-only):
The numerical range  $\chi=(\chi_h,\chi_b]$ corresponds to $z =[z_a,z_b]$. 
At the horizon, $z(\chi \to \chi_h) \to -\infty$ and the potential $U$ approaches a finite value. $\chi$ and $z$ are unbounded outwards. 
Thus the situation is the same with HBH1-II. 
Here, we also introduce $z^*$ in order to set phase constants of different eigenvalue solutions equal.

\clearpage
\begin{figure*}[btph]
\begin{center}
\includegraphics[width=0.3\textwidth]{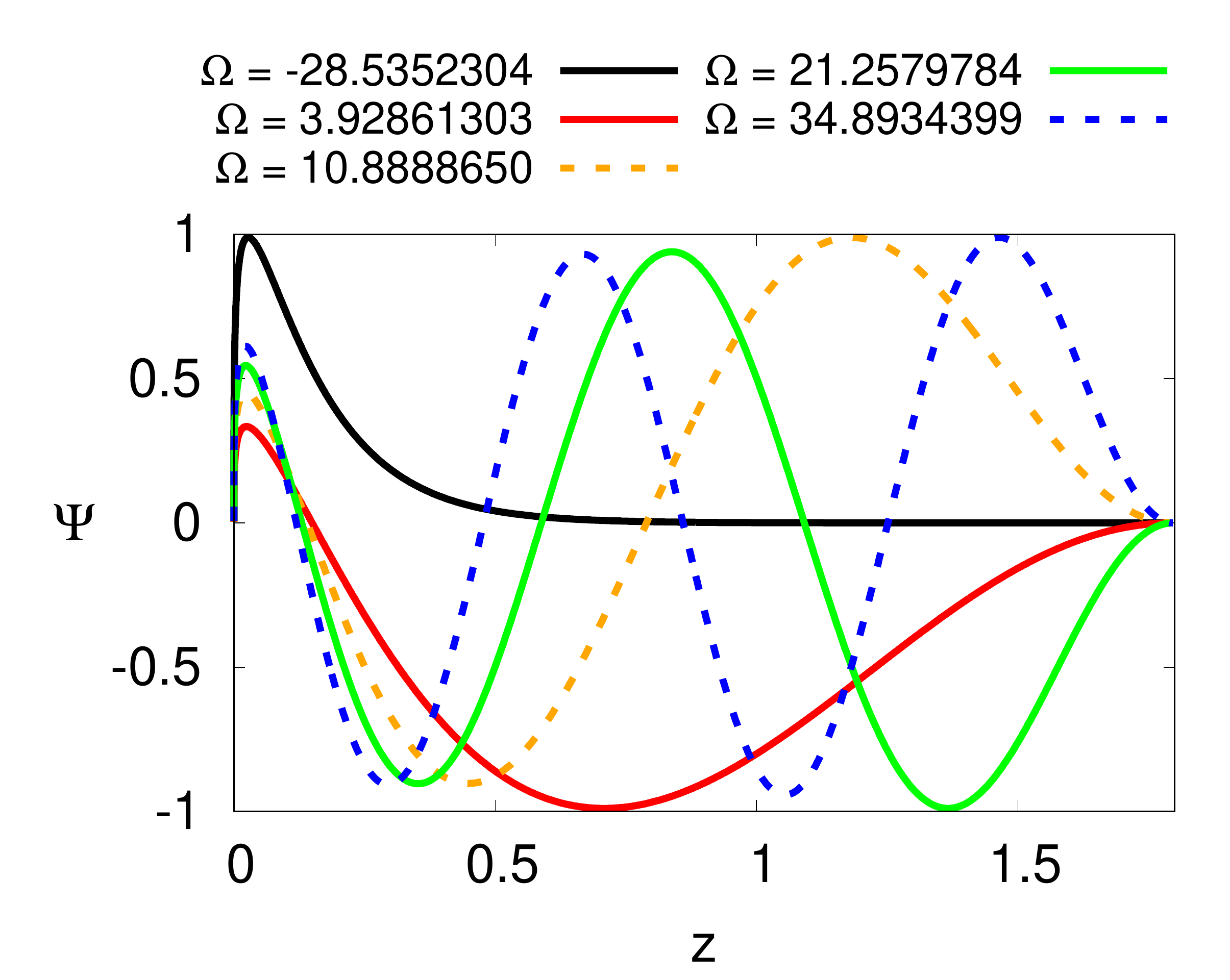}
\includegraphics[width=0.3\textwidth]{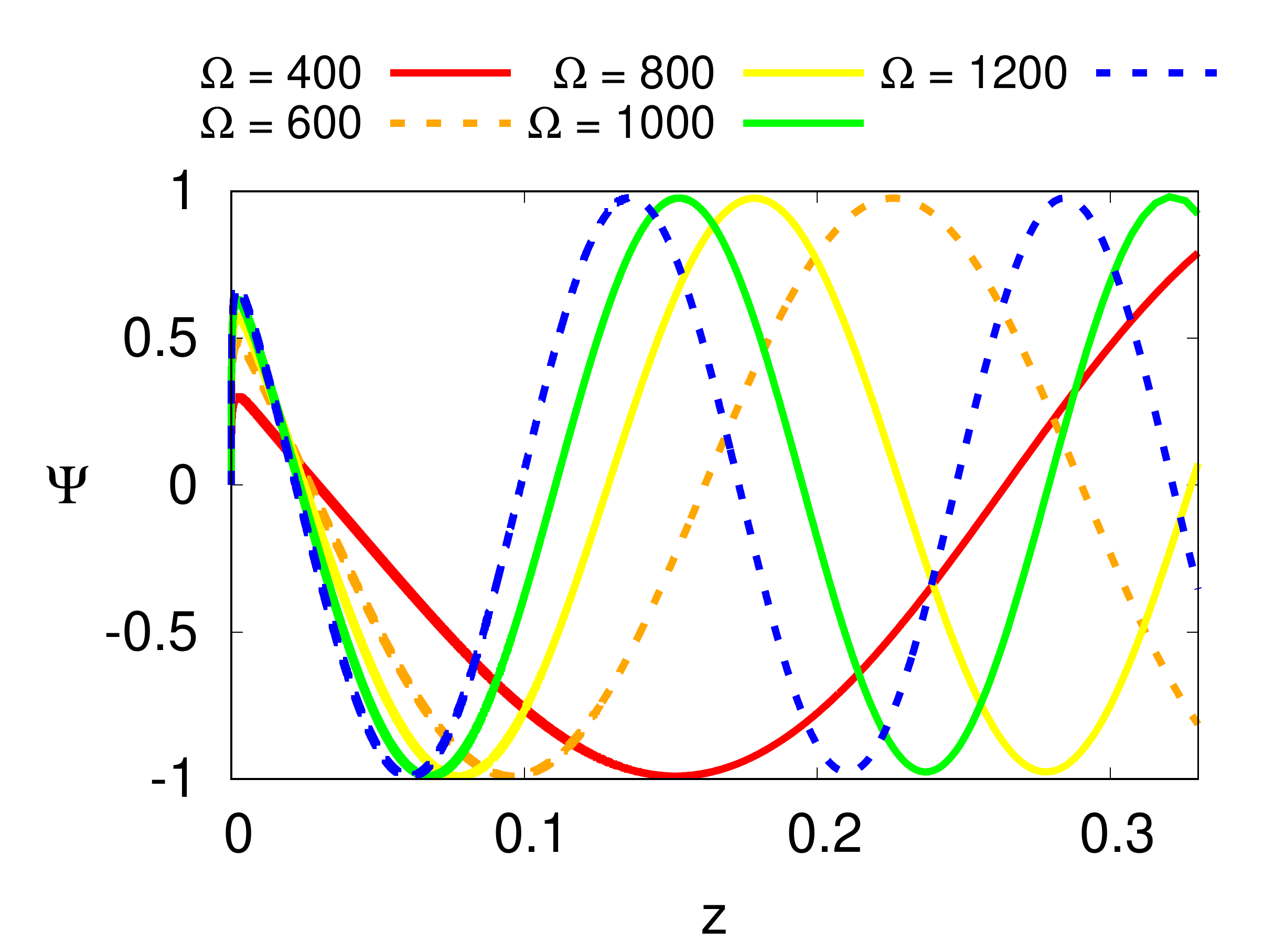}\\
\hspace{0.1in} (a)  SST1 \hspace{1.5in} (b) SBH1-I \\
\includegraphics[width=0.3\textwidth]{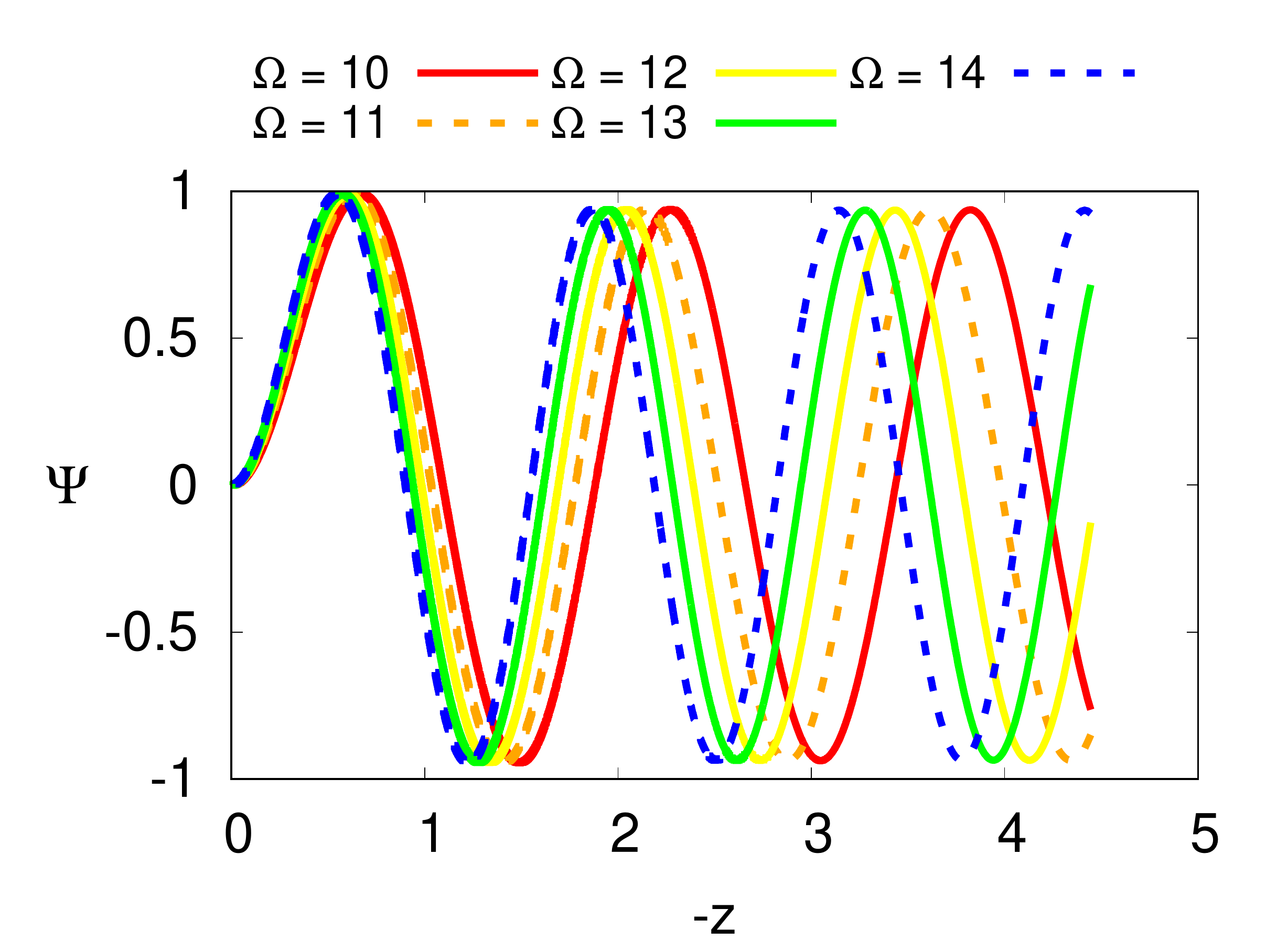}
\includegraphics[width=0.3\textwidth]{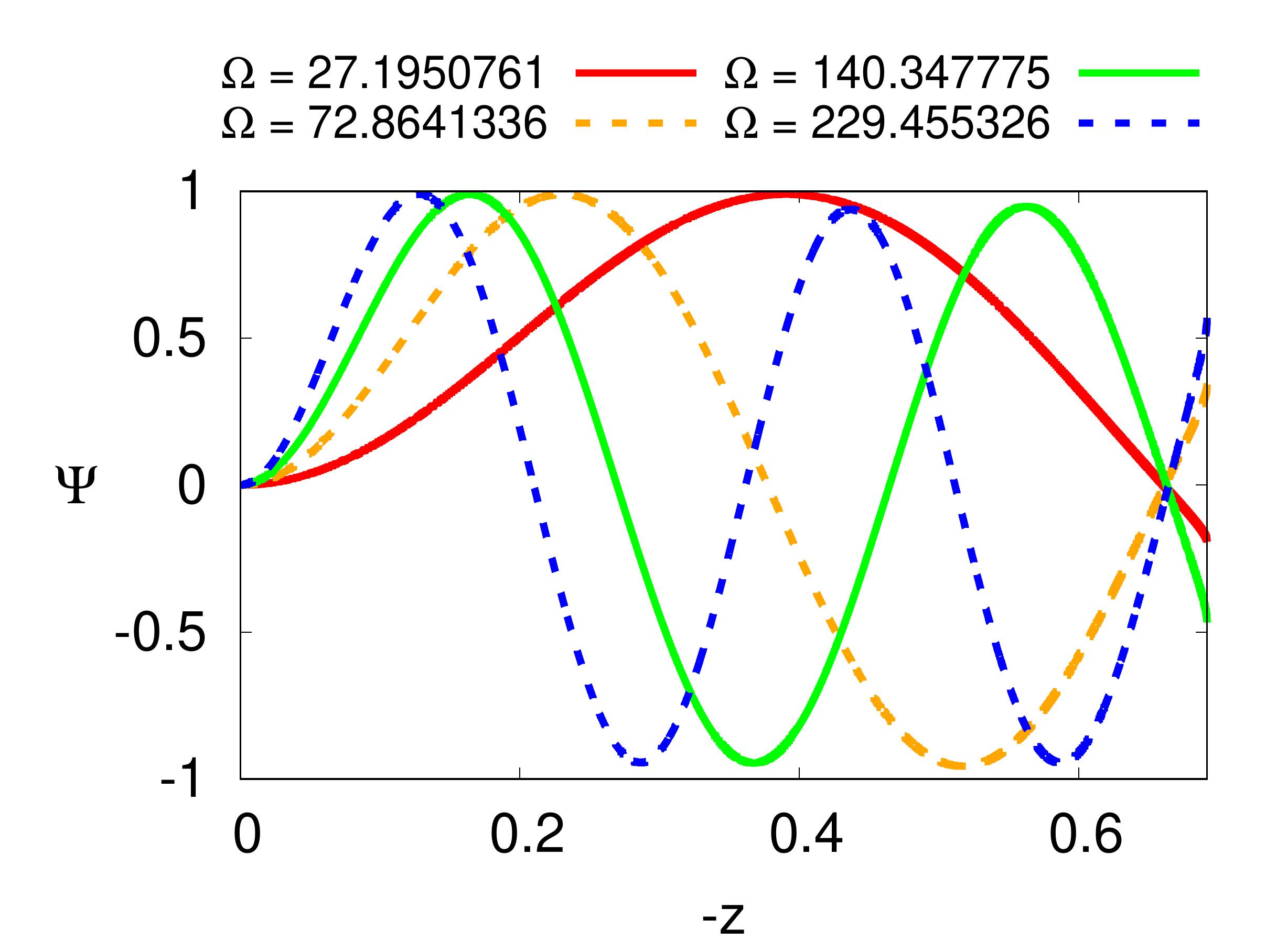}\\
\hspace{0.1in} (c)  SBH1-II \hspace{1.5in} (d) SBH2\\
\end{center}
\caption{Plot of $\Psi(z)$ for $S_3$.
(a) Charged static case for $K=5/9$, $Q=1$, $R_0=1$.
There is no horizon.
Numerical range: $\chi = [0,\pi/2)$ corresponds to $z =[0,z_b)$.
(b) Charged black-hole case for $K=1$, $Q=1$, $R_0=1$ and 
inside the inner horizon. 
Numerical range: $\chi = [0,\chi_-)$ corresponds to $z =[0,\infty)$,
but we take the practical range as
$\chi = [0,\chi_b(<\chi_-)]$ corresponding to $z =[0,z_b]$.
(c) Charged black-hole case for $K=1$, $Q=1$, $R_0=1$ and 
outside the outer horizon.
Numerical range: $\chi = (\chi_+,\pi/2]$ corresponds to $z =(-\infty,0]$ by adjusting $z_0$. Here we changed the signature of z to make the $z$ range as 
$[0,z_b]$.
(d) Fluid-only black-hole case for $K=0.5$.
There is one horizon.
Numerical range: $\chi =(\pi/2,\pi]$ corresponds to $z =(z_a,z_b]$. Here we ajusted $z_0$ and changed the signature of z to make the $z$ range as 
$[0,z_b]$.
}
\end{figure*}
\vspace{12pt}
\begin{figure*}[btph]
\begin{center}
\includegraphics[width=0.3\textwidth]{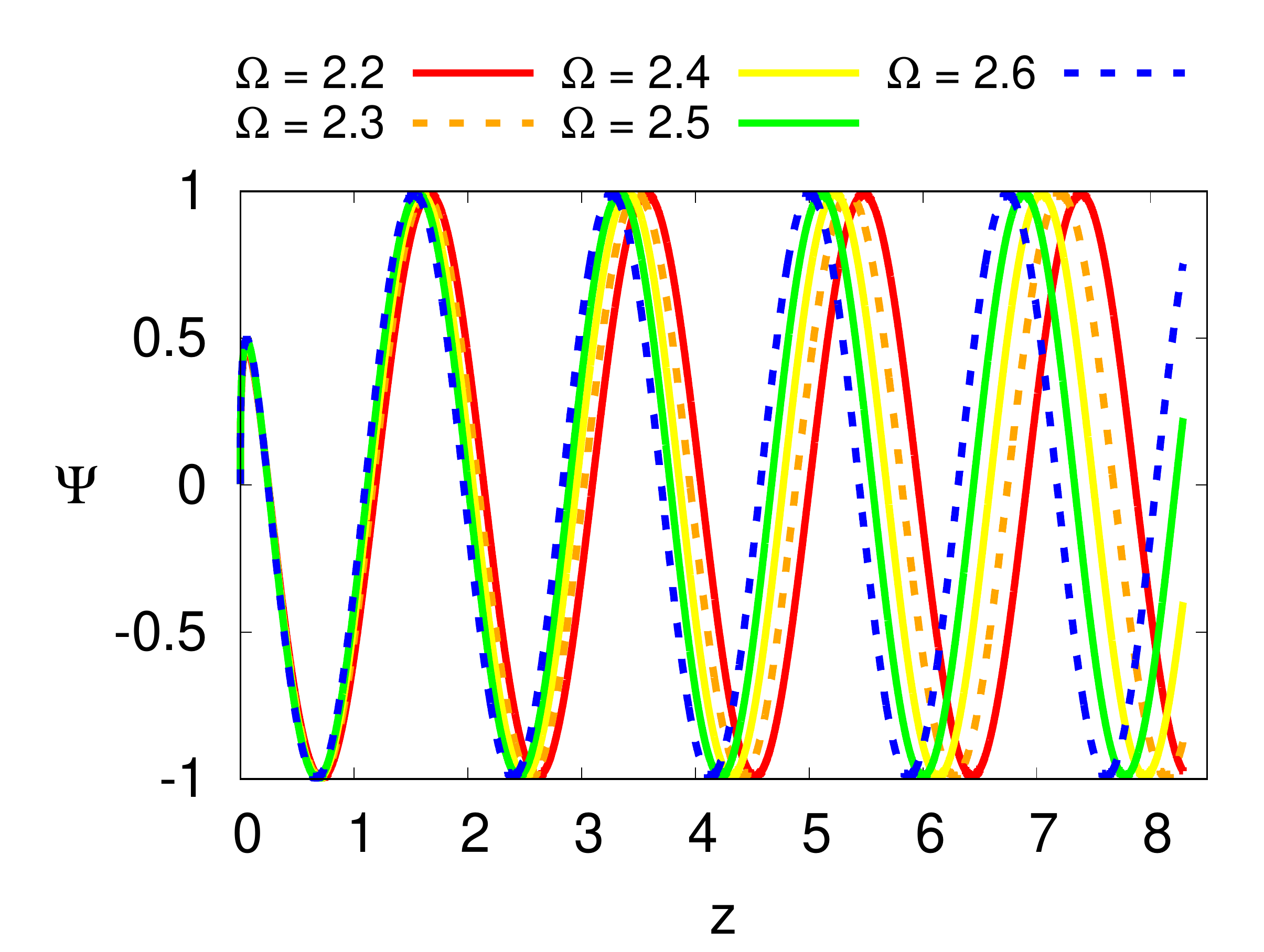}
\includegraphics[width=0.3\textwidth]{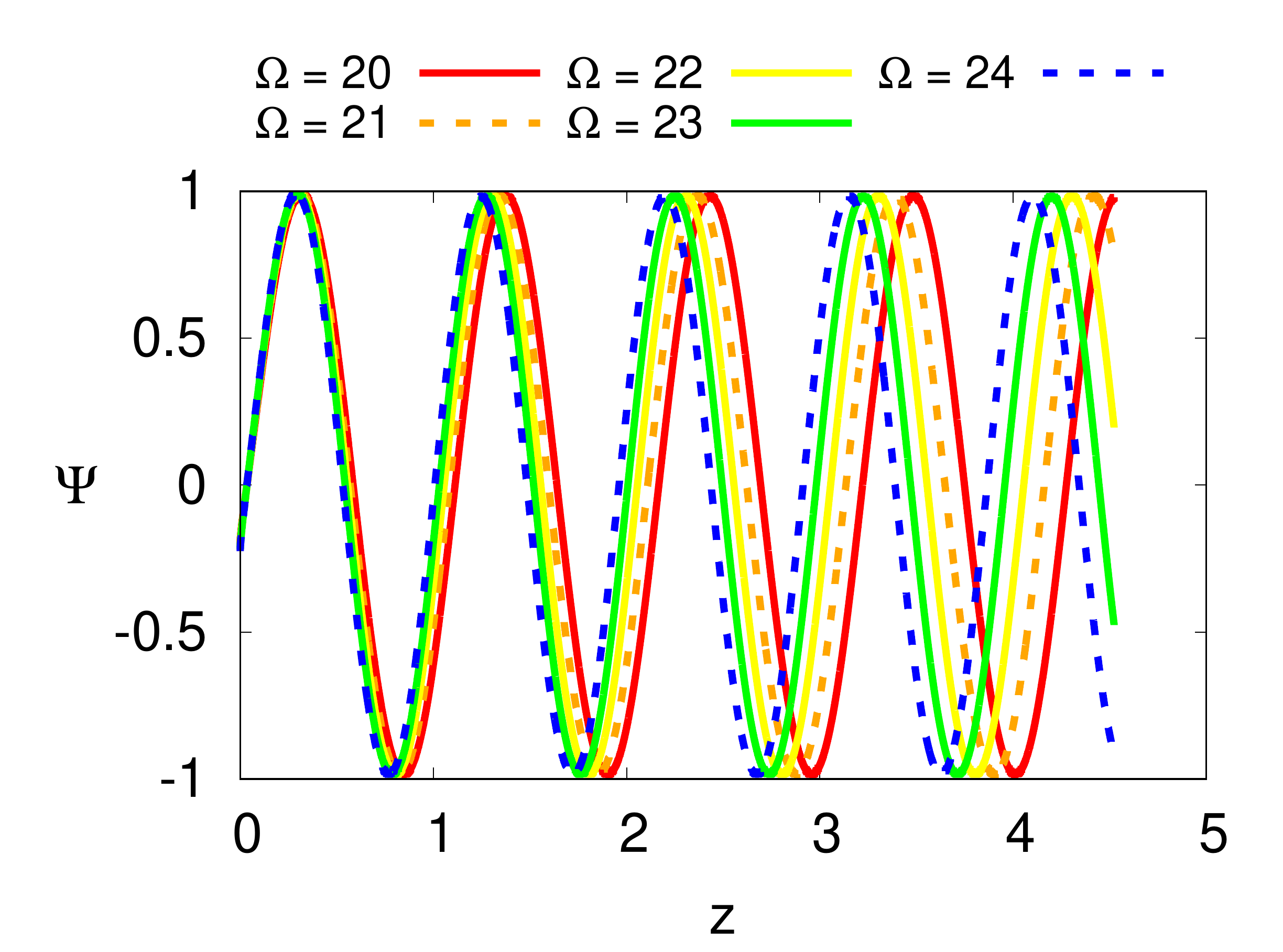}\\
\hspace{0.1in} (a)  HST1 \hspace{1.5in} (b) HST2 \\
\includegraphics[width=0.3\textwidth]{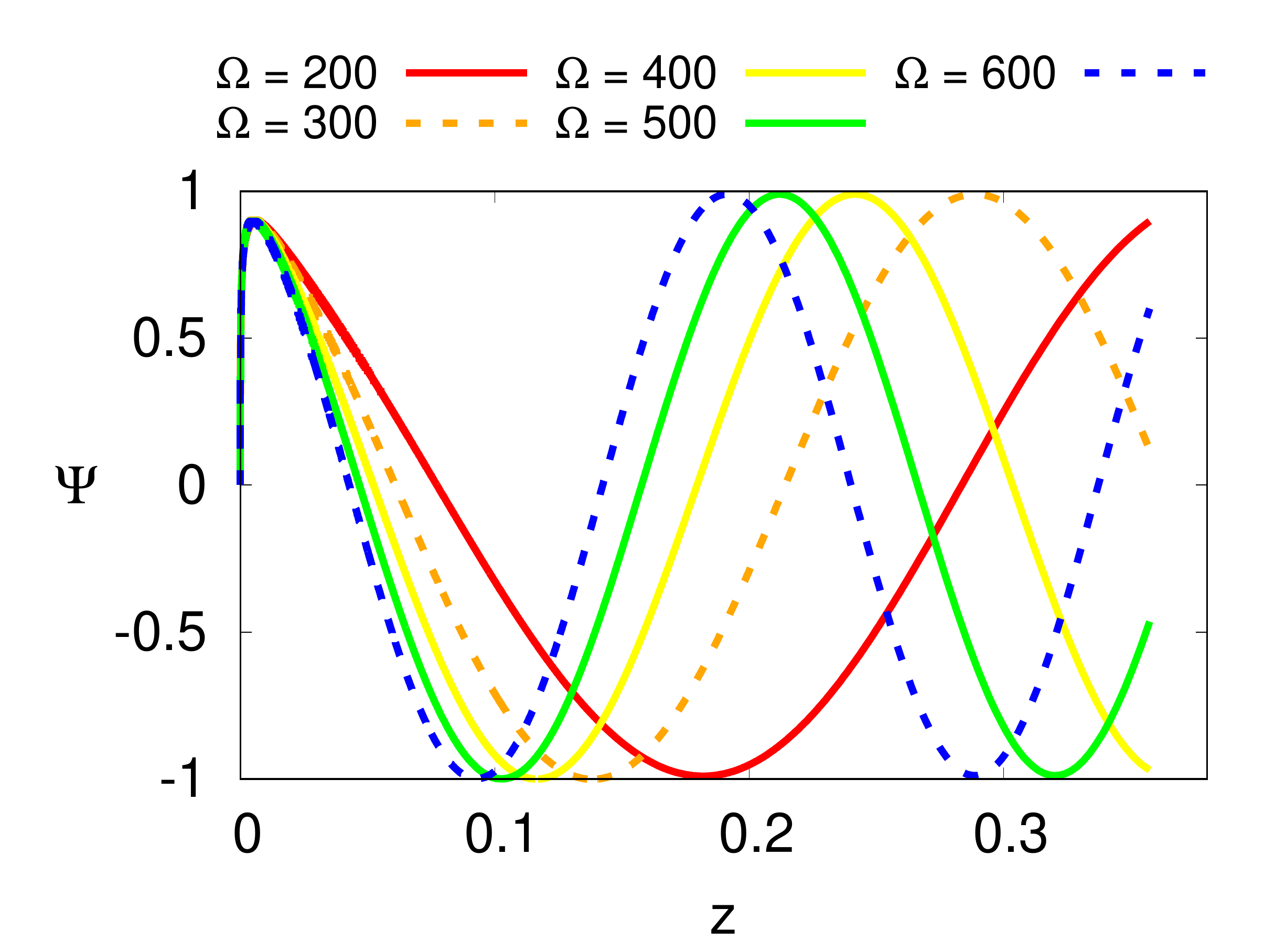}
\includegraphics[width=0.3\textwidth]{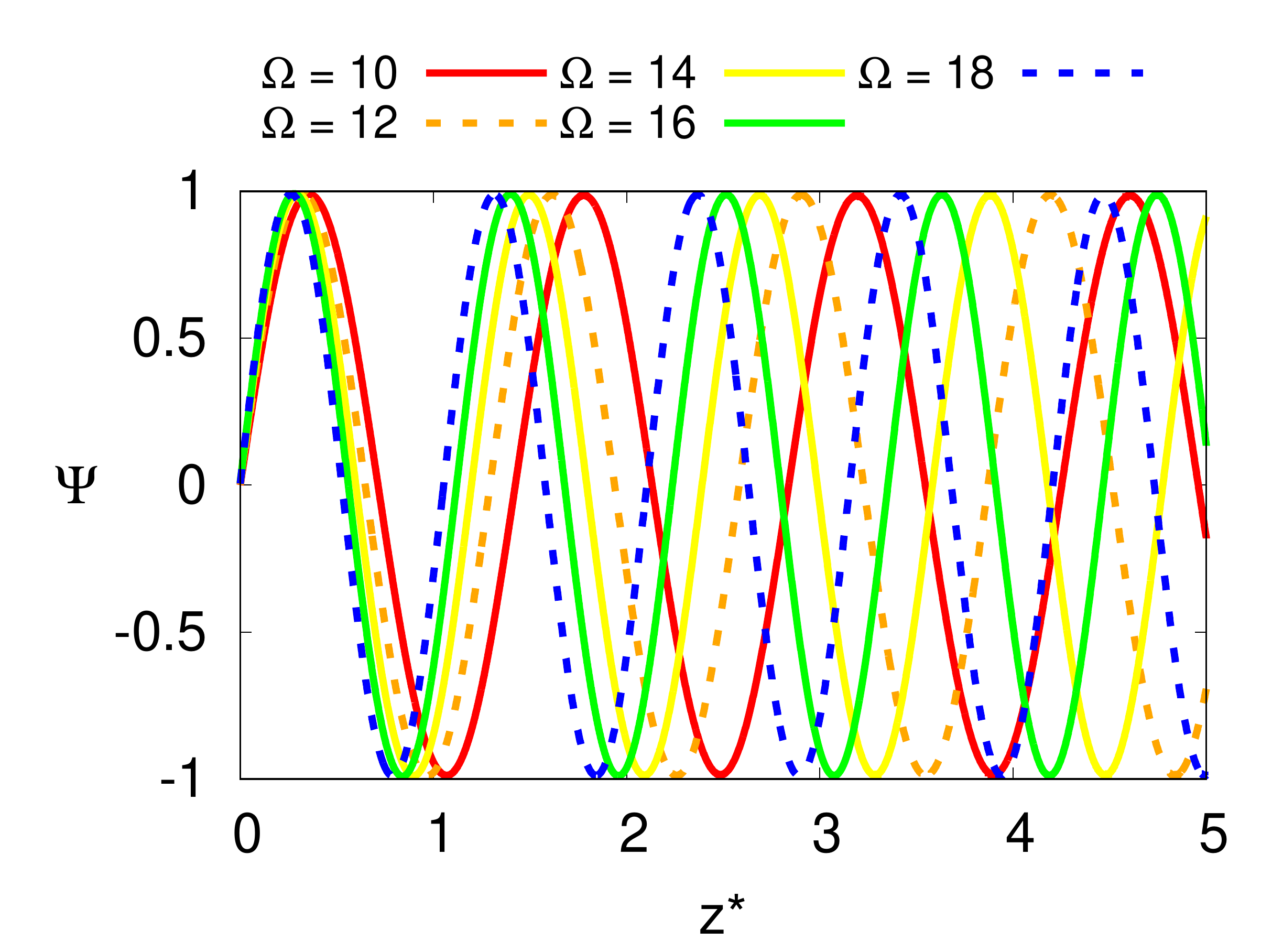}
\includegraphics[width=0.3\textwidth]{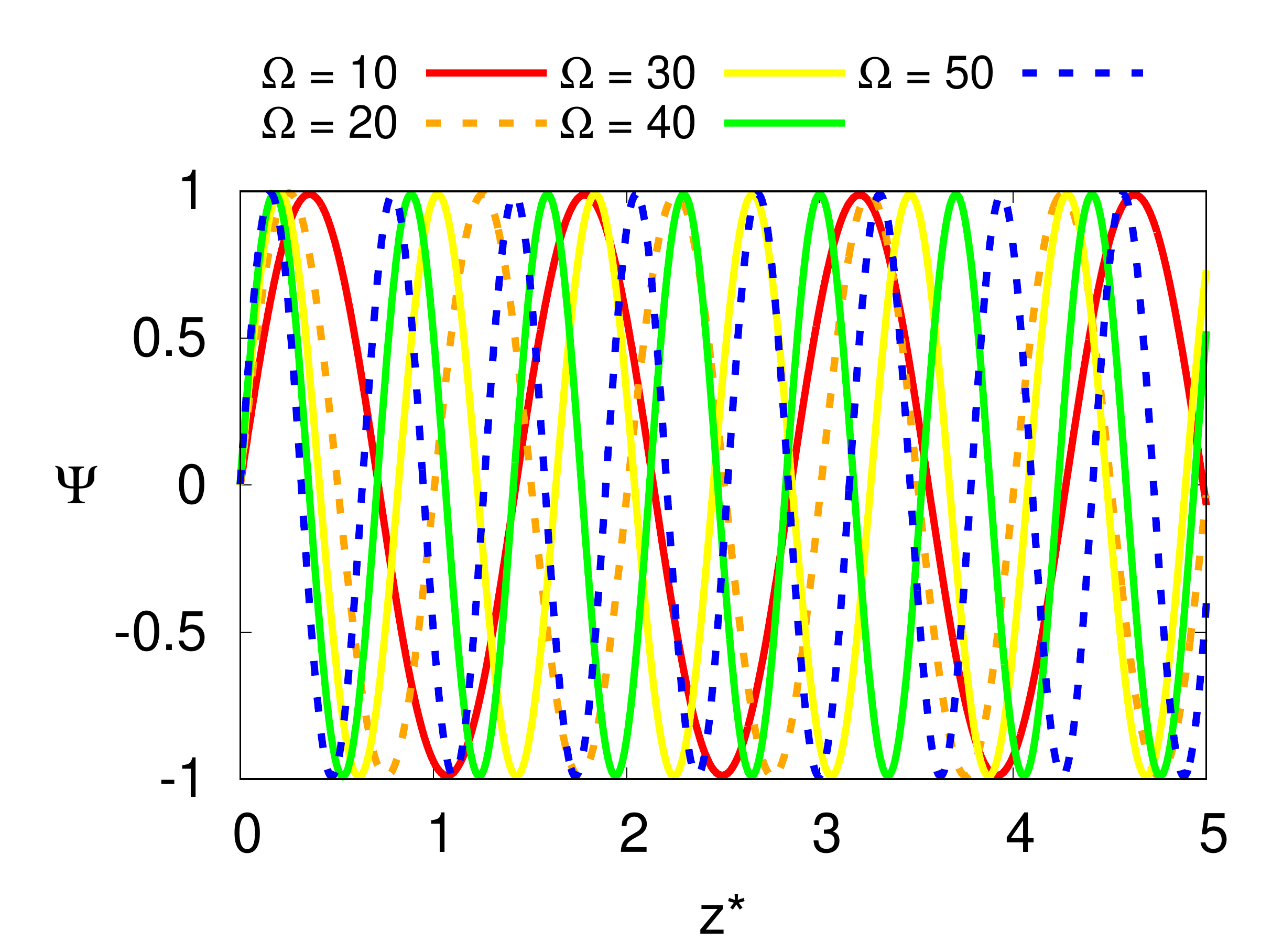}\\
\hspace{0.2in} (c)  HBH1-I \hspace{1.5in} (d) HBH1-II \hspace{1.5in} (e) HBH2 \\
\end{center}
\caption{Plot of $\Psi$ for $H_3$.
(a) Charged static case for $K=0.5$, $Q=1$, $R_0=1$.
There is no horizon.
Numerical range: $\chi = [0,\chi_b]$ corresponds to $z =[0,z_b]$,
where $z_b$ can be arbitrarily large.
(b) Fluid-only static case for $K=-0.5$.
There is no horizon.
Numerical range: $\chi = [0,\chi_b]$ corresponds to $z =[0,z_b]$,
where $z_b$ can be arbitrarily large.
(c) Charged black-hole case for $K=1$, $Q=1$, $R_0=1$ and inside the inner horizon.
Numerical range: $\chi = [0,\chi_-)$ corresponds to $z =[0,\infty)$,
but we take the practical range as
$\chi = [0,\chi_b(<\chi_-)]$ corresponding to $z =[0,z_b]$.
(d) Charged black-hole case for $K=1$, $Q=1$, $R_0=1$ and outside the outer horizon.
Numerical range: $\chi = (\chi_+,\infty)$ corresponds to $z =(-\infty,\infty)$,
but we take the practical range as
$\chi = [\chi_a(<\chi_+),\chi_b]$ corresponding to $z =[z_a,z_b]$. Here we translate $z$ coordinate to set the phase constants of the different eigenvalue solutions equal.
(e) Fluid-only black-hole case for $K=0.5$.
There is one horizon.
Numerical range: $\chi = (\chi_h,\chi_b]$ corresponds to $z =(-\infty,z_b]$,
but we take the practical range as
$\chi = [\chi_a(>\chi_h),\chi_b]$ corresponding to $z =[z_a,z_b]$. Here we translate $z$ coordinate to set the phase constants of the different eigenvalue solutions equal.
}
\end{figure*}

\clearpage
\begin{figure*}[btph]
\begin{center}
\includegraphics[width=0.3\textwidth]{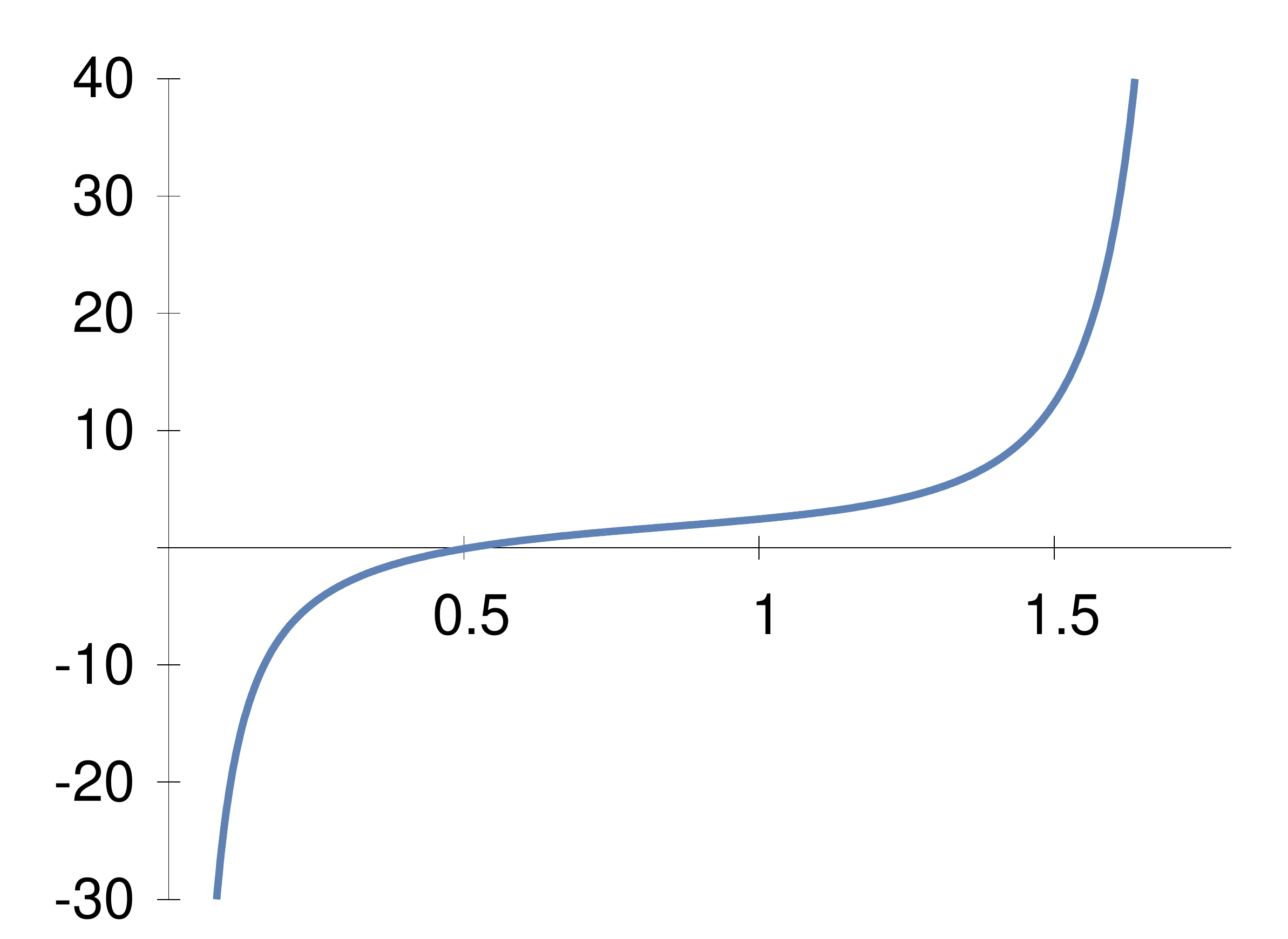}
\includegraphics[width=0.3\textwidth]{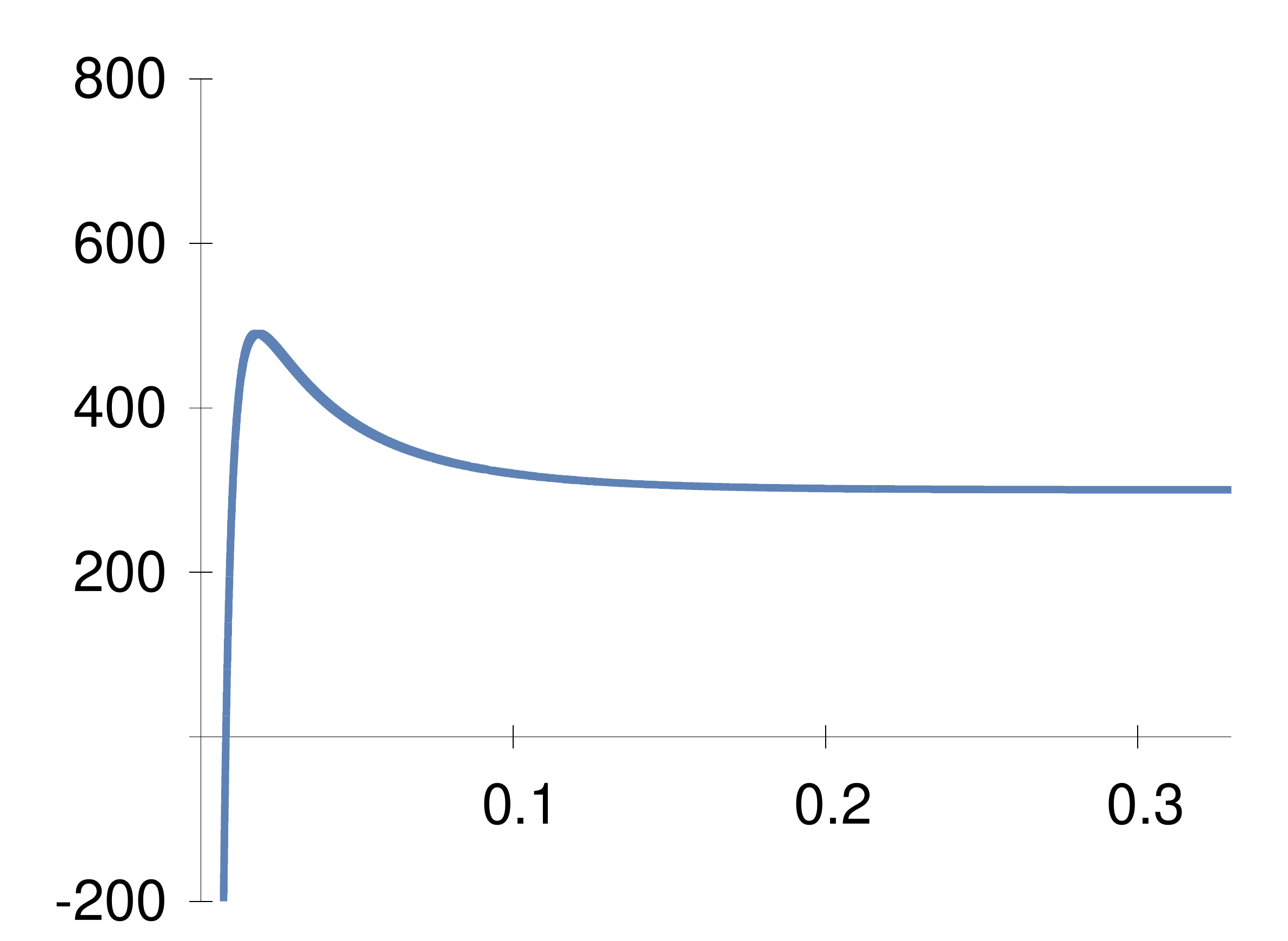}\\
\hspace{0.2in} (a) SST1 \hspace{1.5in} (b) SBH1-I  \\
\includegraphics[width=0.3\textwidth]{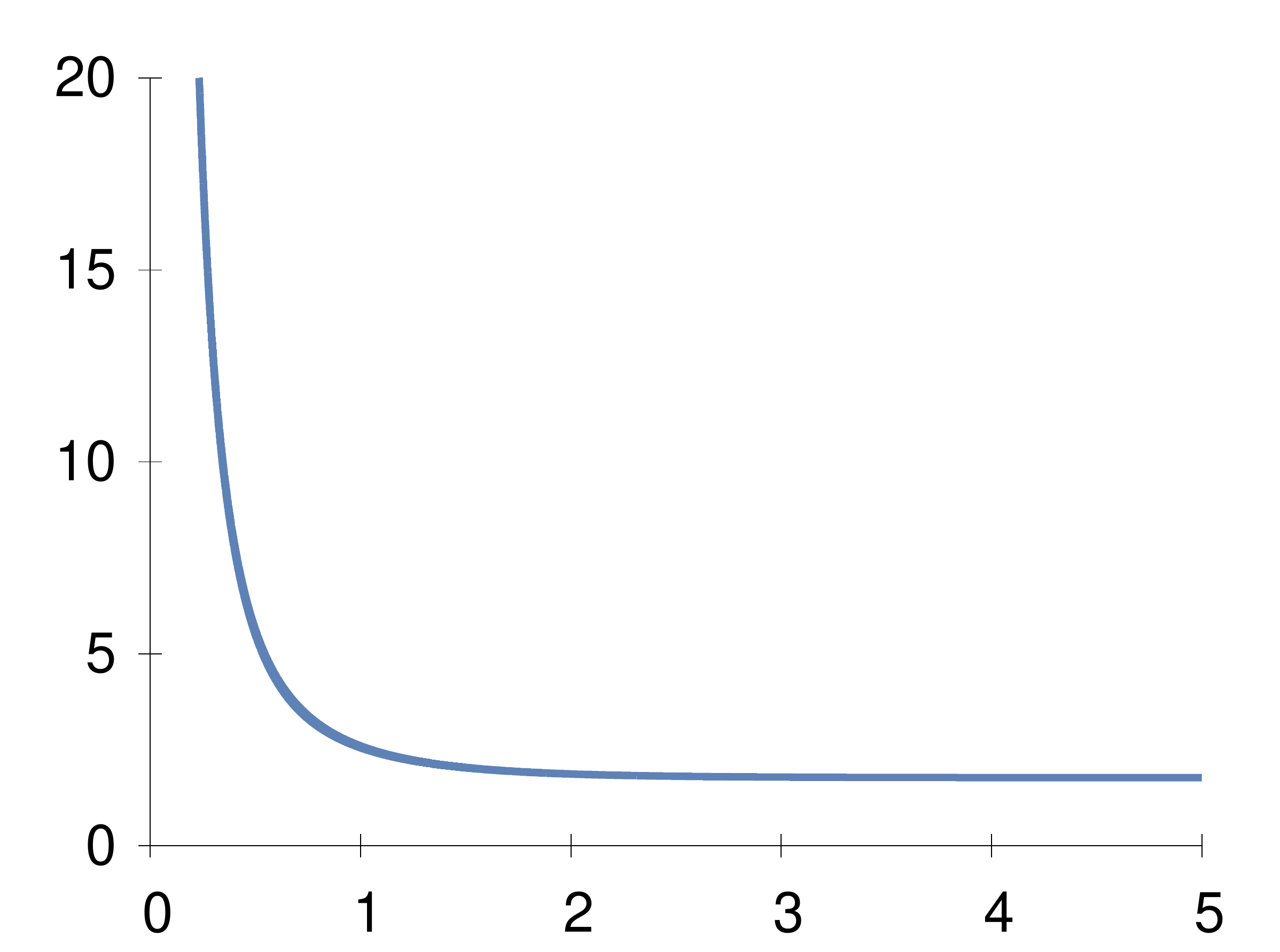}
\includegraphics[width=0.3\textwidth]{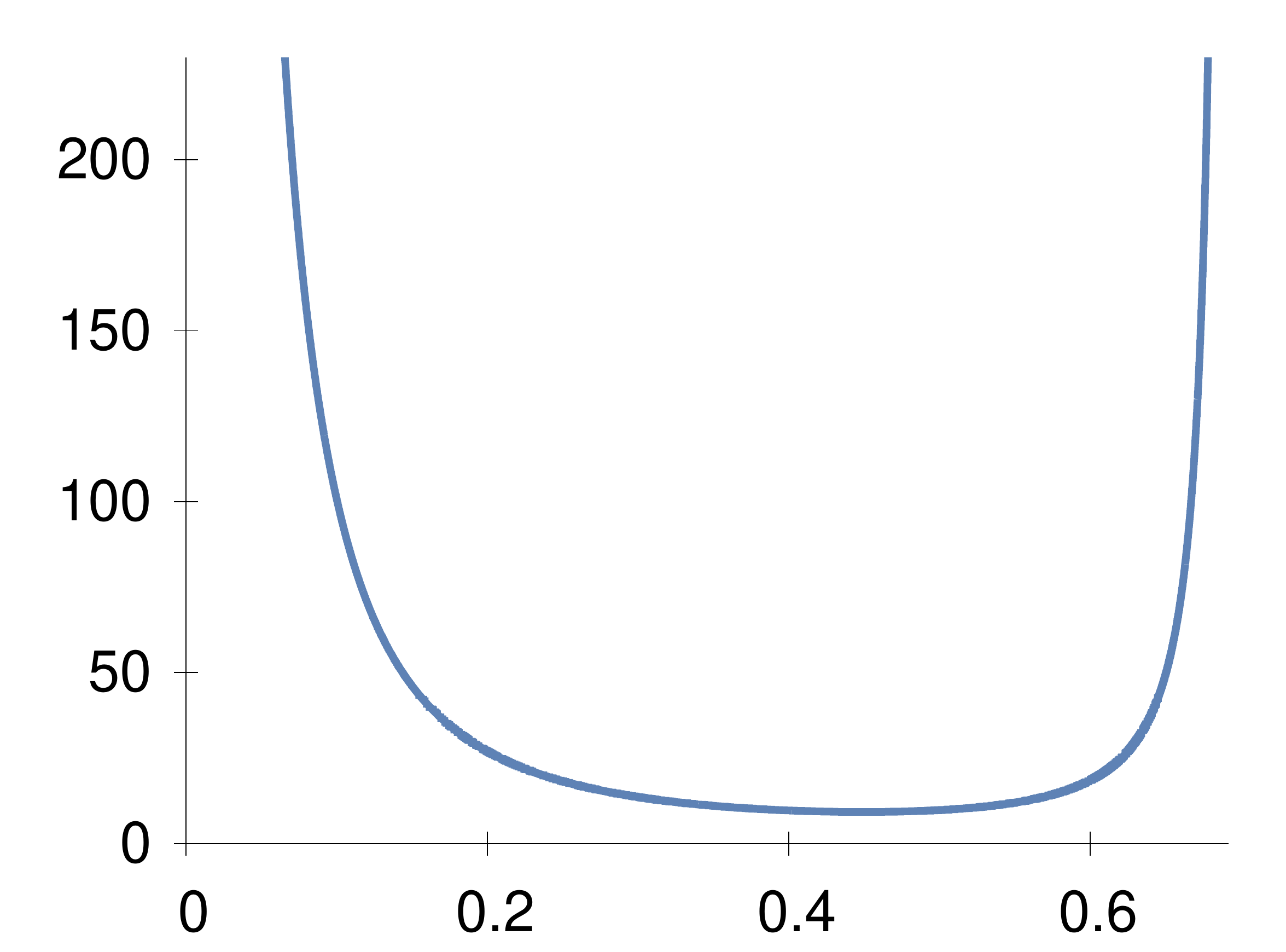}\\
\hspace{0.2in} (c) SBH1-II \hspace{1.5in} (d) SBH2\\
\end{center}
\caption{
Plot of $U(z)$ for $S_3$.
(a) Charged static case for $K=5/9$, $Q=1$, $R_0=1$.
Numerical range: $\chi = [0,\pi/2)$ corresponds to $z =[0,z_b)$.
(b) Charged black-hole case for $K=1$, $Q=1$, $R_0=1$ and 
inside the inner horizon.
Numerical range: $\chi = [0,\chi_-)$ corresponds to $z =[0,\infty)$,
but we take the practical range as
$\chi = [0,\chi_b(<\chi_-)]$ corresponding to $z =[0,z_b]$.
(c) Charged black-hole case for $K=1$, $Q=1$, $R_0=1$ and 
outside the outer horizon.
Numerical range: $\chi = (\chi_+,\pi/2]$ corresponds to $z =(-\infty,0]$ by adjusting $z_0$. Here we changed the signature of z to make the $z$ range as 
$[0,z_b]$.
(d) Fluid-only black-hole case for $K=0.5$.
There is one horizon.
Numerical range: $\chi =(\pi/2,\pi]$ corresponds to $z =(z_a,z_b]$. Here we ajusted $z_0$ and changed the signature of z to make the $z$ range as 
$[0,z_b]$.
}
\end{figure*}
\vspace{12pt}
\begin{figure*}[btph]
\begin{center}
\includegraphics[width=0.3\textwidth]{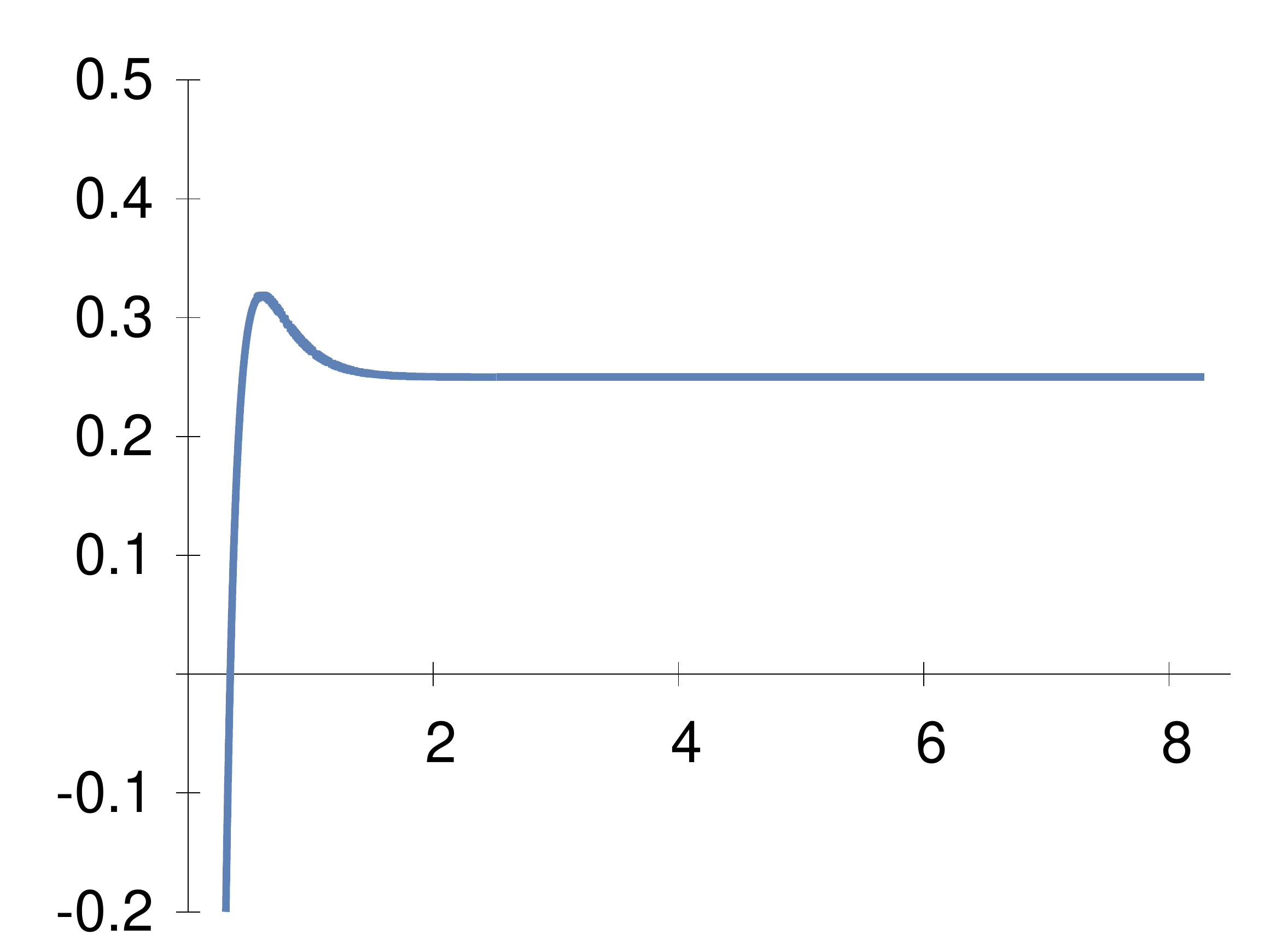}
\includegraphics[width=0.3\textwidth]{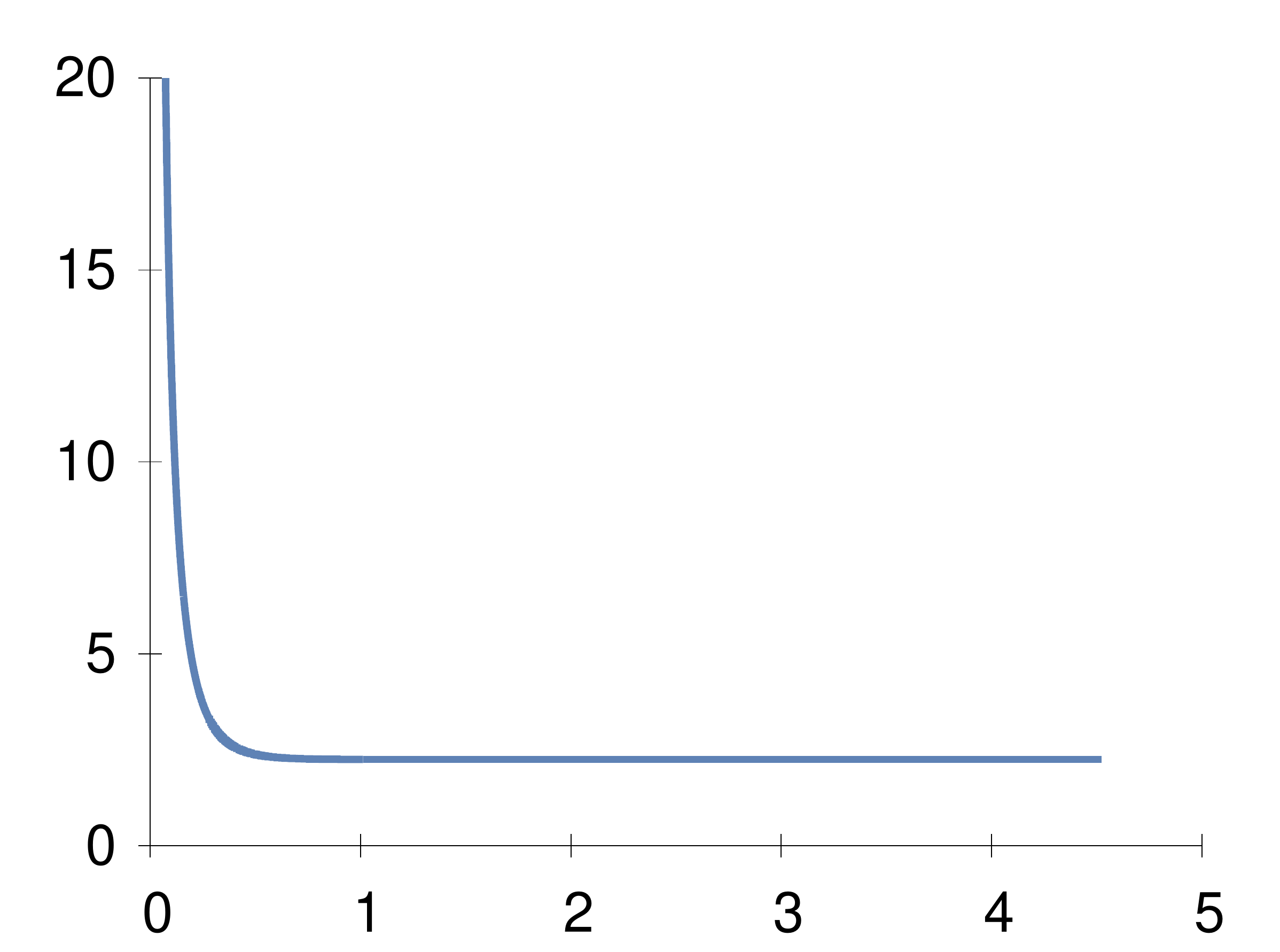}\\
\hspace{0.2in} (a)  HST1 \hspace{1.5in} (b) HST2 \\
\includegraphics[width=0.3\textwidth]{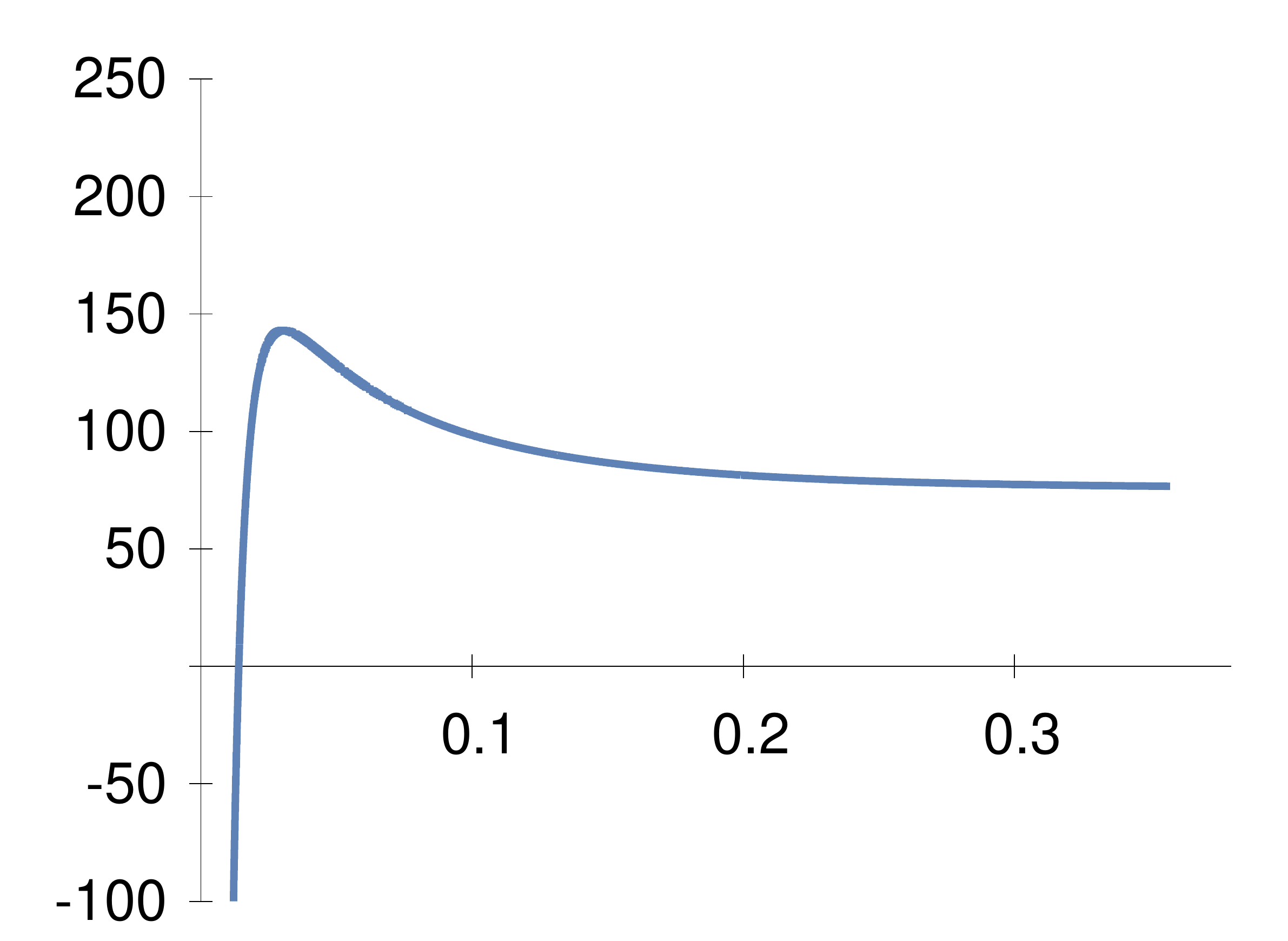}
\includegraphics[width=0.3\textwidth]{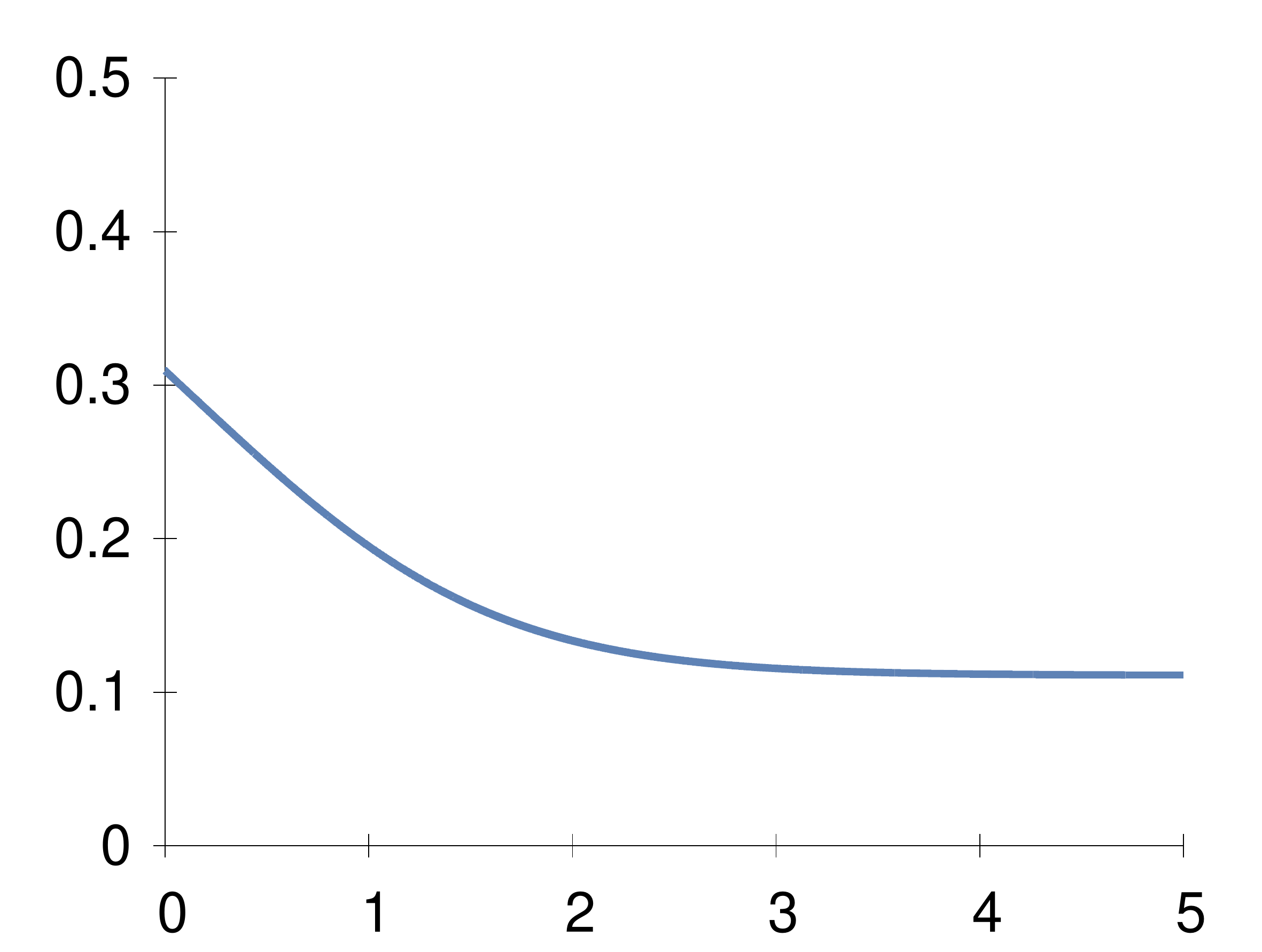}
\includegraphics[width=0.3\textwidth]{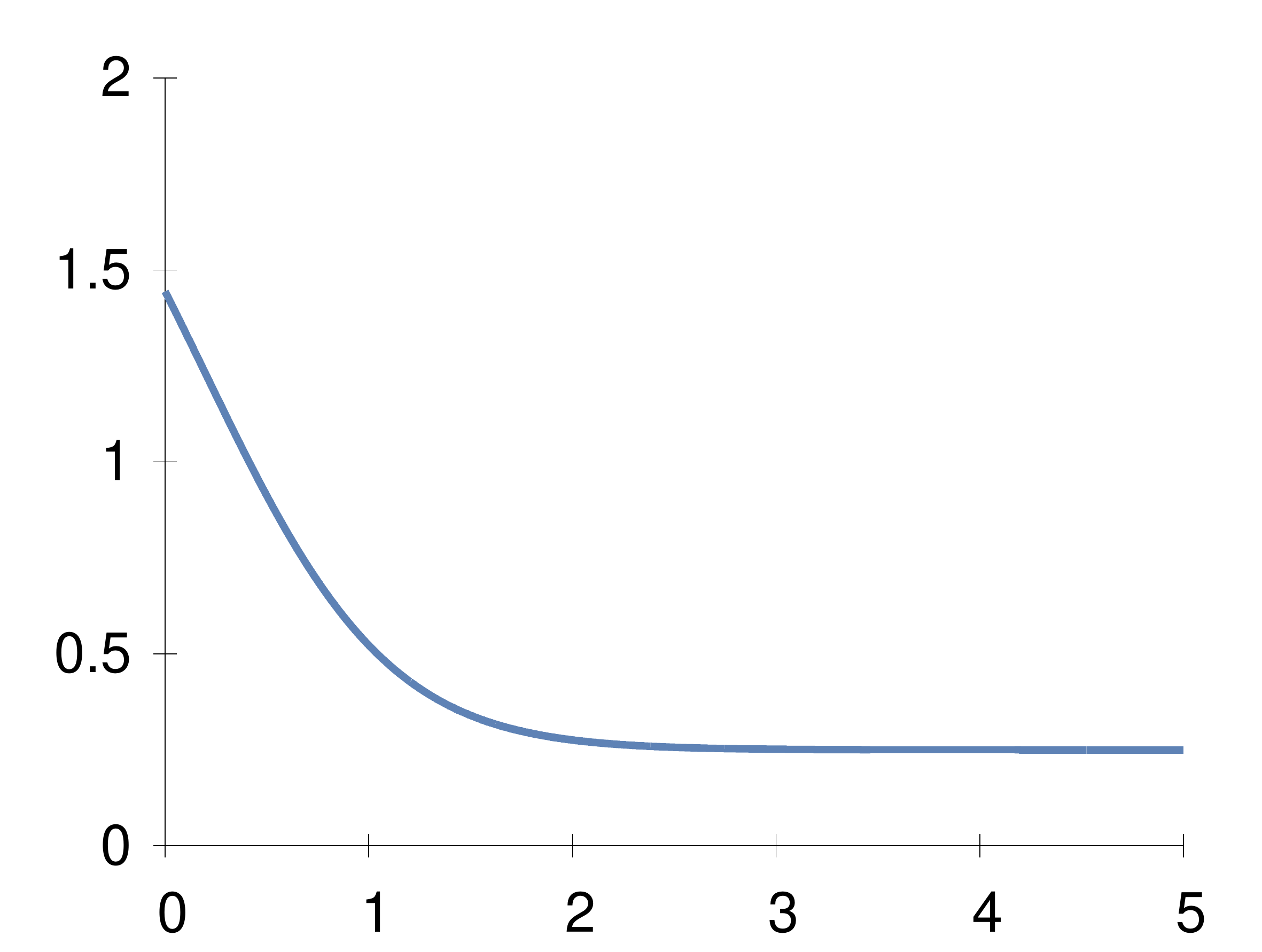}\\
\hspace{0.3in} (c)  HBH1-I \hspace{1.5in} (d) HBH1-II \hspace{1.5in} (e) HBH2 \\
\end{center}
\caption{
Plot of $U(z)$ for $H_3$.
(a) Charged static case for $K=0.5$, $Q=1$, $R_0=1$.
Numerical range: $\chi = [0,\chi_b]$ corresponds to $z =[0,z_b]$.
(b) Fluid-only static case for $K=-0.5$.
Numerical range: $\chi = [0,\chi_b]$ corresponds to $z =[0,z_b]$.
(c) Charged black-hole case for $K=1$, $Q=1$, $R_0=1$ and inside the inner horizon.
Numerical range: $\chi = [0,\chi_-)$ corresponds to $z =[0,\infty)$,
but we take the practical range as
$\chi = [0,\chi_b(<\chi_-)]$ corresponding to $z =[0,z_b]$.
(d) Charged black-hole case for $K=1$, $Q=1$, $R_0=1$ and outside the outer horizon.
Numerical range: $\chi = (\chi_+,\infty)$ corresponds to $z =(-\infty,\infty)$,
but we take the practical range as
$\chi = [\chi_a(<\chi_+),\chi_b]$ corresponding to $z =[z_a,z_b]$. 
(e) Fluid-only black-hole case for $K=0.5$.
Numerical range: $\chi = (\chi_h,\chi_b]$ corresponds to $z =(-\infty,z_b]$,
but we take the practical range as
$\chi = [\chi_a(>\chi_h),\chi_b]$ corresponding to $z =[z_a,z_b]$. 
}
\end{figure*}

\clearpage
\section{Conclusions}
We investigated numerically the stability of the static and the black-hole solutions
of fluid with and without electric field.
It was known that the solutions are unconditionally unstable~\cite{Cho:2016kpf,Cho:2017vhl}.
We performed numerical calculations in order to 
obtain the eigenvalues and the eigenfunctions for unstable modes.
We also tried to get the numerical solutions for the stable modes.

We investigated seven classes as summarized in Tab. II.
For $S_3$ and $H_3$, the static and the black-hole solutions 
have been studied for the charged and the fluid-only cases.
(Only the $S_3$ static fluid-only class does not exist.)

Since we introduced the metric perturbation as Eq.~\eqref{f1g1},
we performed numerical calculations in the region 
where the spacetime is static, 
i.e., where $t$ and $\chi$ are the temporal and the spatial coordinates, respectively. 
We solve the perturbation equation \eqref{PE1} for $\varphi(\chi)$,
but we interpret the solution as an eigenfunction $\Psi(z)$ 
in the Schr\"odinger-type equation \eqref{PE2}.

Even in the static region, we limit the numerical region by the boundary
where the Schr\"odinger potential $U$ diverges.
This applies to two classes (SST1, SBH2),
for which the numerical range becomes compact in the new radial coordinate $z$.
For these classes, the numerical results show 
that the eigenvalues are discrete. 
The eigenfunctions exhibit  the increase in the number of the node
as it is expected.

For the charged black-hole classes (SBH1, HBH1),
we performed numerical calculations at two static regions;
inside the inner horizon and outside the outer horizon.
Although the numerical range in $\chi$ is finite,
the range in the Schr\"odinger-type coordinate $z$ is infinite.
For the rest of $H_3$ classes (HST1, HST2, HBH2),
the space in $\chi$ is unbounded and in $z$ as well.
For these five classes, the corresponding potential approaches a constant
asymptotically as $z$ increases.  
Therefore, the eigenfunctions look like plane waves 
and the eigenvalues become a continuum.

We also tried to check if there exists a stable mode.
For the discrete spectrum of SST1 [Fig. 3(a)],
we could observe the ground state is the stable mode with $\Omega = -\omega^2/\sigma <0$.
This eigenfunction possesses no node between the boundaries,
while the eigenfunctions for the excited (unstable) modes with $\Omega >0$ 
exhibit the increasing number of nodes.
For the discrete spectrum of SBH2 [Fig. 3(c)],
The eigenfunction of the lowest unstable mode that we found exhibits a node.
We conclude that this mode represents the ground state
since the right boundary value for this class is nonzero
as shown in Eq.~\eqref{SBH2bc}.

For the rest of classes, the eigenfunctions that we found are plane waves
with the eigenvalue larger than the asymptotic value of the  potential $U$.
We tried to search the solution with the eigenvalue 
smaller than the asymptotic value of the  potential $U$,
but we could not observe a conspicuous figure of the wave function
which can be regarded as an eigenfunction.

In this work, we introduced the linear spherical scalar perturbations.
Introducing more complicated perturbations such as anisotropic perturbations,
would be interesting.
The result of this work is very useful in studying the stability
of the monopole with fluid, which is under investigation.
The geometry of a global monopole described by a scalar-field triplet 
in the presence of fluid exhibits
the characters of both the monopole and the fluid.
For the static monopole-fluid configuration, 
we can consider the stability.
We expect that the instability of fluid may be controlled by the scalar field,
so the fluid-monopole becomes possibly stable.
However, there is a difficulty in studying the stability
because for the monopole-fluid system,
the perturbation fields are not decoupled to give a single
master equation such as Eq.~\eqref{PE1}.
Instead, there are two coupled equations with two perturbation fields.
In this case, the equations cannot be cast into a Schr\"odinger type
as Eq.~\eqref{PE2},
so one cannot obtain the potential $U$.
Then it is hard to imagine how the wave functions will behave.
In principle, the numerical calculation always gives the solution
for given boundary conditions unless there are any divergences
in the wave function.
Once one gets the numerical results for given boundary conditions,
it is hard to judge if the results are the ones under the physical considerations.
In this case, the numerical results for fluid in this work
will be very useful in selecting the proper solutions among various numerical results.
If one turns down the monopole scale to a small scale
by taking a small value of the symmetry-breaking scale of the monopole,
one can make the effect of fluid more dominant.
The solution for this case must be close to the solution that we obtained in this work.

\acknowledgements
Authors are grateful to Hyeong-Chan Kim and Gungwon Kang for useful discussions.
This work was supported by the grants from the National Research Foundation
funded by the Korean government, No. NRF-2017R1A2B4010738
and No. NRF-2020R1A2C1013266.

\end{document}